\documentclass[11pt,preprint]{aastex}
\usepackage{amsmath,pifont}
\newcommand{\au}{\,{\rm AU}}

\newcommand{\xmark}{\ding{55}}%
\newcommand{\cmark}{\ding{51}}%
\begin{document}
\title{Analysis of terrestrial planet formation by the Grand Tack model: System architecture and tack location}
\author{R.~Brasser\altaffilmark{1}, S. Matsumura\altaffilmark{2}, S. Ida\altaffilmark{1}, S. J. Mojzsis\altaffilmark{3,4}, S. C.
Werner\altaffilmark{5}}
\altaffiltext{1}{Earth-Life Science Institute, Tokyo Institute of Technology, Meguro-ku, Tokyo 152-8550, Japan}
\altaffiltext{2}{School of Science and Engineering, Division of Physics, Fulton Building, University of Dundee, Dundee DD1 4HN, UK}
\altaffiltext{3}{Collaborative for Research in Origins (CRiO), Department of Geological Sciences, University of Colorado, UCB 399, 
2200 Colorado Avenue, Boulder, Colorado 80309-0399, USA}
\altaffiltext{4}{Institute for Geological and Geochemical Research, Research Center for Astronomy and Earth Sciences, Hungarian 
Academy of Sciences, 45 Buda\"{o}rsi Street, H-1112 Budapest, Hungary}
\altaffiltext{5}{The Centre for Earth Evolution and Dynamics, University of Oslo, Sem Saelandsvei 24, 0371 Oslo, Norway}
\begin{abstract}
The Grand Tack model of terrestrial planet formation has emerged in recent years as the premier scenario used to account for 
several observed features of the inner solar system. It relies on early migration of the giant planets to gravitationally sculpt and 
mix the planetesimal disc down to $\sim$1~AU, after which the terrestrial planets accrete from material left in a narrow 
circum-solar annulus. Here we have investigated how the model fares under a range of initial conditions and migration 
course-change (`tack') locations. We have run a large number of N-body simulations with a tack location of 1.5~AU and 2~AU and tested 
initial conditions using equal mass planetary embryos and a semi-analytical approach to oligarchic growth. We make use of a recent 
model of the protosolar disc that takes account of viscous heating, include the full effect of type 1 migration, and employ a 
realistic mass-radius relation for the growing terrestrial planets. Results show that the canonical tack location of Jupiter at 1.5~AU 
is inconsistent with the most massive planet residing at 1~AU at greater than 95\% confidence. This favours a tack farther out at 
2~AU for the disc model and parameters employed. Of the different initial conditions, we find that the oligarchic case is 
capable of statistically reproducing the orbital architecture and mass distribution of the terrestrial planets, while the equal mass 
embryo case is not. 
\end{abstract}
\section{Introduction}
\label{sec:int}
A successful physical model for the formation of the terrestrial planets is a long-standing problem \citep{KT15}. The first 
physically plausible idea came from Safronov (1969) who suggested the earliest stage of accumulation of dust into larger bodies was 
caused by gravitational instability in a thin dust layer. Safronov (1969) showed that relative velocities between bodies are of the 
order of their escape velocity, so the largest body's gravitational cross section is limited by the geometrical one, limiting growth. 
These findings were later used by \cite{W80}, who showed the terrestrial planets coagulated from planetesimals, and that the 
formation of the these planets was linked with the evolution of the asteroid belt. \cite{WS89} elaborated that in a disc of 
planetesimals some would undergo runaway growth and form a sequence of planetary embryos. These embryos would then further collide to 
form the terrestrial planets.\\

These ideas were first rigorously tested by \cite{ki96} who performed numerical simulations of a self-gravitating disc of 
planetesimals. They discovered that some objects in the disc underwent runaway growth, as was predicted, which resulted in a mixed 
population of protoplanets and planetesimals \citep{ki98}. The protoplanets underwent so-called oligarchic growth: all would be 
roughly equally spaced and be of similar mass as each vied for supremacy in accreting the last remaining planetesimals. The 
protoplanets (also dubbed `planetary embryos') subsequently collide to form the terrestrial planets \citep{C01}.\\

Early simulations of terrestrial planet formation yielded estimates for a growth time scale of several tens of millions of years and 
overall results that showed the final terrestrial system would be assembled by 100~Myr \citep{C01}. Most of these early 
simulated systems, however, were found to suffer from an excess eccentricity and inclination of the final planets, but the inclusion 
of a large number of planetesimals to exert dynamical friction alleviated this concern \citep{OB06}. A further chronic and fundamental 
shortcoming with earlier simulations was that the output systematically yielded a far too massive Mars analogue. This predicament led 
\cite{R09} to investigate how the mass of Mars might depend on the orbital configuration of the giant planets. It was found that only 
the current spacing of the gas giants led to the capability of the model to produce a Mars analogue much less massive than Earth, but 
under the special condition that the eccentricities of the gas giants were higher than their current values. As such, \cite{R09} 
highlighted the unrealistic nature of the initial conditions required to explain Mars' low mass, and left the problem as a lingering 
impasse to be solved later.\\

A potential solution presented itself in the work of \cite{H09}, who studied terrestrial planet formation with planetary embryos 
situated in a narrow annulus between 0.7~AU and 1~AU from the Sun. These initial conditions nicely reproduced the mass-semimajor axis 
relationship we have today, with two relatively large terrestrial-type planets book-ended by two much less massive ones. The main 
drawback of that study was that no mechanism was presented to gravitationally truncate the outer edge of the solid disc near 
1~AU. The same could be said for the inner edge, so that no mechanism existed to create such a high-density, narrow annulus with 
which to explain the terrestrial worlds.\\

This quandary led \cite{W11} to propose the so-called Grand Tack scenario, wherein the early gas-driven coupled migration of 
Jupiter and Saturn sculpts the planetesimal disc and truncates it near 1~AU. The Grand Tack at least partially explains 
the formation of a high density region in the inner disc, although it cannot explain the existence of an inner cavity inside of 
roughly 0.7~AU. The inclusion of this scenario leads to a broad outline of how the early solar system evolved: First, 
Jupiter is assumed to form before Saturn, clear the gas in an annulus the width of which is comparable to its Hill radius, and undergo 
inward Type 2 migration \citep{LP86}. The inward migration of Jupiter shepherds material towards the inner portion of the disc while 
also scattering other material outwards, to create an enhanced density region for terrestrial planet formation and that mixes
planetesimals from the innermost and outer portions of the protoplanetary disc. Once Saturn grows to about half its current mass 
\citep{MC09}, it is assumed to partially clear the disc in its vicinity, migrate rapidly at first to catch up with Jupiter, and 
subsequently get trapped in a mean-motion resonance near Jupiter, presumably the 2:3 \citep{ms01,PR11} but it may also have been the 
2:1 \citep{P14}. In this process, these two giant planets clear the disc together. The torque from the interaction with the 
disc is stronger for a shorter separation between the planet and the disc edge. Since Jupiter and Saturn are interacting and Jupiter 
is more massive, it is reasonable to think that Saturn is pushed outwards by Jupiter's perturbation, and the separation from the 
disc edge is smaller for Saturn than for Jupiter. Thus the torque on Saturn can be larger in spite of lower mass. The interaction with 
Jupiter prevents Saturn from creating a cleared annulus in the disc and gas from the outer disc flows past Saturn into the inner disc. 
If the gap-crossing disc gas flow is large enough, the Jupiter-Saturn pair can migrate outwards \citep{ms01,PN08}. Consequently, the 
planets reverse their migration: they ’tack’ as a sail boat would change its direction by steering into and through the wind. Once the 
giant planets have completed this early migration phase, have left the inner solar system and settled in the vicinity of their present 
positions, terrestrial planet formation could proceed as before, but only (as advocated by \cite{H09}) from material in a 
narrow circum-solar annulus. In this manner \cite{W11} successfully reproduced the mass-semimajor axis distribution of 
the inner planets if the reversal of Jupiter occurred at 1.5~AU because they truncated the inner edge of their planetesimal disc at 
0.5~AU. A successful feature of their model is that it also accounts for the apparent compositional differences across the asteroid 
belt \citep{DC14}.\\

It is worth noting, however, that this step wise reconstruction of the early evolution of the planetary system has some pitfalls of 
varying severity. For example, the particular configuration and outward migration of the giant planets favoured by the Grand Tack is 
only supported for a narrow set of initial conditions \citep{AM12} and is not universal \citep{ZZ10}. There may also be other pathways 
to produce such a high density region through a deficit of material near Mars \citep{I14}, although this idea was recently undermined 
in a follow-up study \citep{I15}. Lastly, up to this point the Grand Tack fails to reproduce the current mass and location of Mercury, 
most likely because dynamical models always truncate the disc near 0.5~AU or beyond. Clearly further study is needed in both the 
gas-driven evolution of the giant planets and the subsequent formation and evolution of the terrestrial planets to explain what we see 
in our own solar system.\\

With this in mind, we sought to scrutinise the Grand Tack model and its consequences over most of the age of the solar system, by 
running a large number of Grand Tack simulations with a range of initial conditions and varying tack locations. Our work also include d
several dynamical effects that have hitherto been ignored. We report on the various methods that were employed to quantify whether or 
not Grand Tack successfully reproduces the observed dynamical features of the inner solar system and whether one set of initial 
conditions and tack location is more favourable than another.\\

This report is organised as follows. In Section~2 we introduce several additions to the original Grand Tack simulations of \cite{W11} 
and justify our choice of disc model and the inclusion of type 1 migration. Section~\ref{sec:nm} describes our initial conditions, 
while Section~\ref{sec:methods} describes our numerical methods. Section~\ref{sec:success} explains our criteria for a set of 
simulations to successfully reproduce the current observed dynamical properties of the inner solar system. This is followed by 
Sections~\ref{sec:results15} and Section~\ref{sec:results20} where we describe the results of our numerical simulations. 
Section~\ref{sec:dis} is reserved for a discussion, and we draw our conclusions in Section~\ref{sec:conc}.

\section{Deviations from the original Grand Tack model}
\label{sec:dev}
In addition to the simple reproduction of the Grand Tack scenario, for this study we also chose to employ a substantially 
different model for the protoplanetary disc than that used by \cite{W11}. An explanation for this choice is provided below. We have 
also included the effect of type 1 embryo migration. 

\subsection{Protoplanetary disc}
\begin{figure}[t]
\resizebox{\hsize}{!}{\includegraphics[]{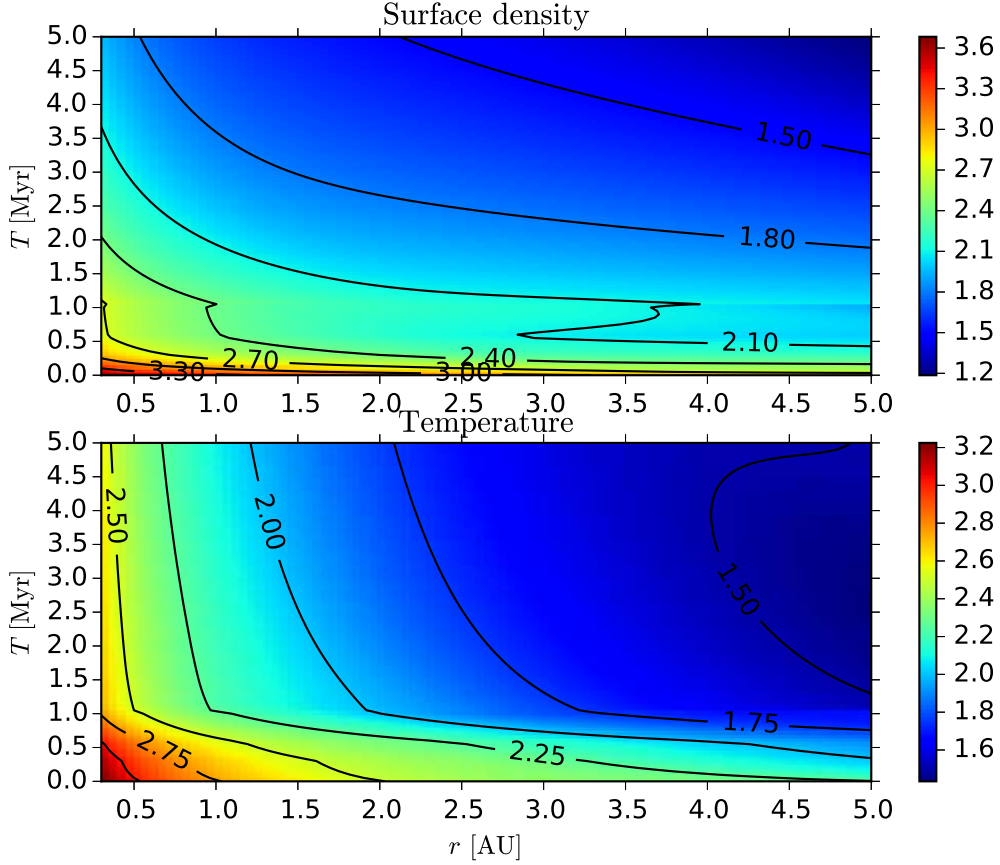}}
\caption{Contour plots of $\log (\Sigma/1\,{\rm g\,cm}^{-2})$ (top) and $\log (T/1\,{\rm K})$ (bottom) as a function of distance to 
the Sun (horizontal axis) and age in Myr (vertical axis).}
\label{fig:bdd}
\end{figure}

\cite{W11} employed the protoplanetary disc model of \cite{mc07}, which in turn was based on the work of \cite{GH06}. The surface 
density of their disc profile is of the form $\Sigma(r) =\Sigma_0 \exp(-r^2/R^2)$, where $R\sim 200\,\au$ is a scaling constant. This 
Gaussian profile of the surface density is markedly different from the oft-employed power law slopes found elsewhere in the 
literature e.g. \cite{H98}. The scaling constant at 1~AU is $\Sigma_0=100$~g~cm$^{-2}$, which is much lower than the usual value 
of 1700--2400~g~cm$^{-2}$ \citep{H81}. Since the disc model of \cite{mc07} is not widely used and relies on a 
constant viscosity, $\nu$, rather than a constant $\alpha$-viscosity, we decided to make use of the disc model of \cite{B14}, which is 
based on the study by \cite{H98}. \cite{B14} give fitting formulae to compute the disc's surface density, temperature and scale height 
as a function of both heliocentric distance and time. Initially the disc's gas surface density at 1~AU is 2272~g~cm$^{-2}$ and the 
temperature is 576~K in the midplane so that the scale height at 1~AU is about 0.057~AU and the metallicity is equal to the solar 
value. This is higher than most traditional models have assumed and the higher temperature is caused by viscous heating.
These fitting formulae are valid, however, as long as any embedded planet is not massive enough to significantly alter the disc 
structure, such as opening an annulus, and as long as the disc remains mostly unperturbed. This latter requirement may not be entirely 
true because of the proximity of Jupiter, whose presence imposes a change in the disc's surface density and temperature. For a 
radiative disc, this change in the temperature profile will result in a change in the disc surface density, but for a viscous disc, 
which is the region we work in, this effect is less severe. Besides, a much lower disc temperature, caused by the influence of 
Jupiter, would mean that the ice line is close to 1~AU very early on, which is inconsistent with solar system formation \citep{M15}. 
Here we adopt the disc of \cite{B14} and set the $\alpha$ viscosity equal to 0.005, use a solar metallicity and a molecular weight of 
the gas of 2.3 amu \citep{B14}. The gas surface density scales as $\Sigma(r) \propto r^{-\alpha}$ and the temperature profile is $T(r) 
\propto r^{-\beta}$, where $\alpha=1/2$ and $\beta=6/7$. These power law relations are accurate out to $\sim$5~AU, which is the region 
we are interested in, so we adopted these profiles throughout the disc. In Fig.~\ref{fig:bdd} we plotted the evolution of the surface 
density (top) and the temperature bottom as a function of time and distance to the Sun. There is a very rapid decay for the first 
1~Myr and a slower decay after that. After 5~Myr we photo-evaporate the disc away over the next 100~kyr. 

\begin{figure}[t]
\resizebox{\hsize}{!}{\includegraphics[]{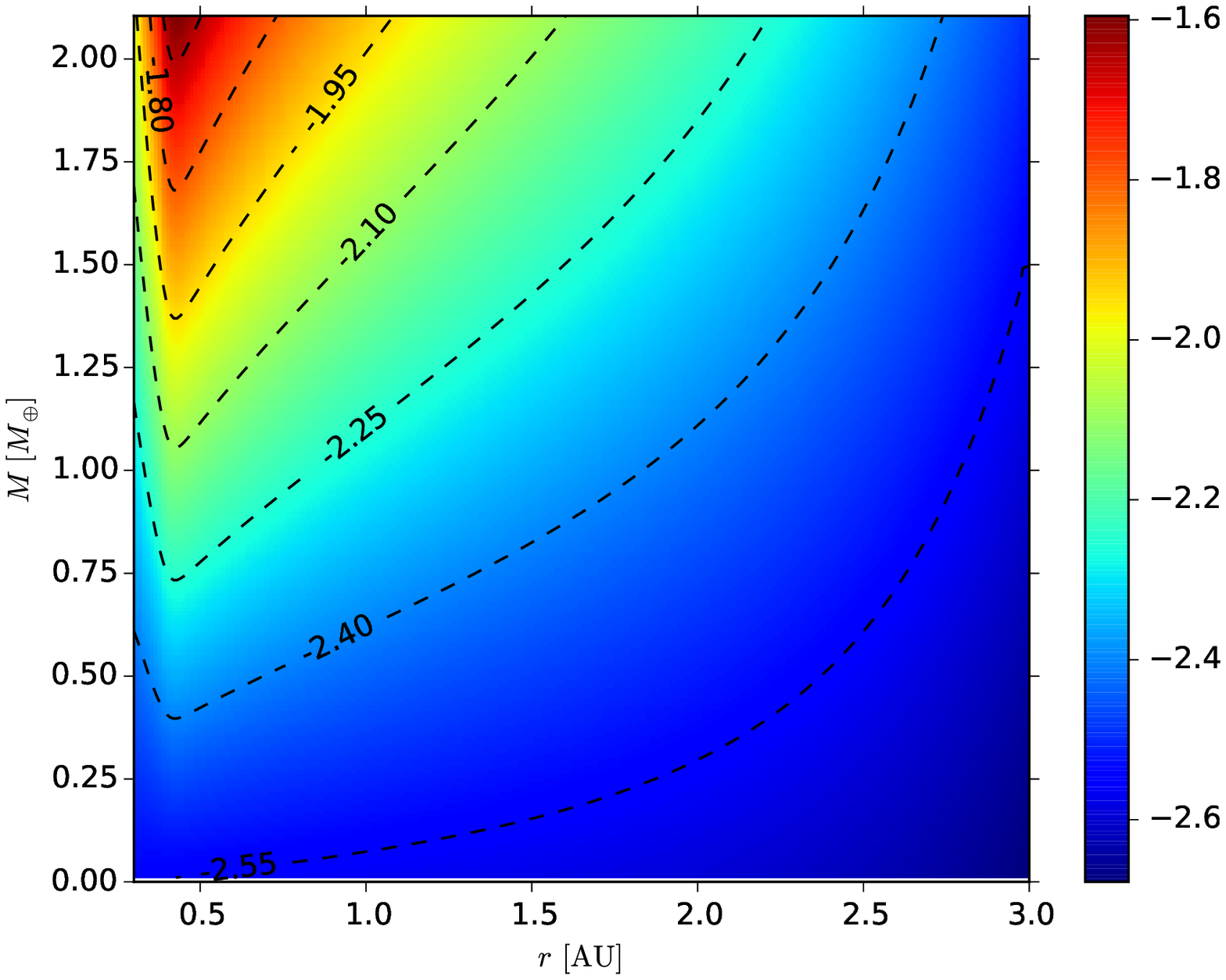}}
\caption{Contour plots of the normalised torque on a planetary embryo as a function of semi-major axis and mass.}
\label{fig:torqueue}
\end{figure}

\subsection{Embryo migration}
Although \cite{W11} decided not to include the effect of type 1 migration \citep{tanaka02} on the planetary embryos, we decided to 
take it into account to determine whether (or not) it drastically affects the outcome of the simulations. For the migration 
prescription we follow \cite{cn08}, which is partially based on the work of \cite{tanaka02}. We chose not to employ the 
non-isothermal approach of \cite{P11} because generally none of the terrestrial planets are massive enough to begin outward migration 
apart from at the very late stages when the disc surface density is low. Thus, for simplicity, we shall adhere to the isothermal case. 
The specific decelerations experienced by the planetary embryos due to the disc on the eccentricity, inclination and semi-major axis 
are given by
\begin{eqnarray}
\vec{a}_e &=& -2\frac{(\vec{v}\cdot\vec{r})}{r^2t_e}\vec{r}, \\
\vec{a}_i &=& -\frac{v_z}{t_i}\vec{k}, \\
\vec{a}_m &=& -\frac{\vec{v}}{t_m},
\label{eq:mig}
\end{eqnarray}
where $\vec{r}$ and $\vec{v}$ are the position and velocity vectors of the embryo, $\vec{k}$ is the unit vector in the 
$z$-direction, $v_z$ is the $z$-component of the velocity, and $t_e$, $t_i$ and $t_m$ are the time scales to damp the eccentricity, 
inclination and semi-major axis. The latter quantities depend in a complicated manner on the semi-major axis, eccentricity, 
inclination, surface density and temperature of the gas, and mass of the embryo. We refer the interested reader to \cite{cn08} 
and \cite{tanaka02} for details and for conversion to the cartesian frame. All of the above time scales are a function of the wave 
time, given by 
\citep{TW04}
\begin{eqnarray}
t_{\rm wav} &=& \Bigl(\frac{M_\odot}{m_{\rm emb}}\Bigr)\Bigl(\frac{M_\odot}{\Sigma 
r^2}\Bigr)\Bigl(\frac{c}{r\Omega_{\rm K}}\Bigr)^4\Omega_{\rm K}^{-1}, \nonumber \\
&=&\Bigl(\frac{M_\odot}{m_{\rm emb}}\Bigr)\Bigl(\frac{M_\odot}{\Sigma 
r^2}\Bigr)\Bigl(\frac{H}{r}\Bigr)^4\Bigl(\frac{r}{v_{\rm K}}\Bigr)
\end{eqnarray}
where we used $H/r=c/v_{\rm K}$, $\Omega_{\rm K}$ is the orbital frequency, $c^2=\gamma k_{\rm B}T/\mu m_{\rm p}$ is the sound speed, 
$v_{\rm K}$ is the orbital speed and $H$ is the scale height, $\gamma$ is the ratio of specific heats (taken as 7/5), $k_{\rm B}$ is 
the Boltzmann constant, $m_{\rm p}$ is the mass of the proton and $\mu$ is the molecular weight of the gas, assumed to be 2.3 amu. The 
value of $t_{\rm wav}$ depends sensitively on the slopes of the surface density and temperature profiles but the product $m_{\rm 
emb}t_{\rm wav} \propto a^{3/2-2\beta+\alpha}$ is much less sensitive and is what determines the migration rate of the embryos. We 
generally have $m_{\rm emb}t_{\rm wav} \sim a^{2/7}$ to $a^2$ for our disc profile and the minimum disc profile of \cite{H81} 
(which has $\alpha=3/2$ and $\beta=1/2$).\\

\cite{W11} included the first two deceleration contributions in equations (1-3) since these are tidal effects by the disc 
that damp the eccentricity and inclination, but they omitted the third (equation 3) which is in part responsible for the inward 
migration of the planetary embryos. For the disc that we have chosen we compute near 1~AU $t_m \sim 2\,(0.1~M_\oplus/m_{\rm 
emb})(2000~{\rm g}\,{\rm cm}^{-2}/\Sigma)$~Myr, so any inward migration that the embryos experience will likely be severely restricted 
because the migration time scale is longer than the disc lifetime and increases as the disc surface density decreases. However $t_e 
~\sim 21$~kyr for a Mars-sized body near 1~AU so the damping effect is a lot stronger than the migration. For the disc employed by 
\cite{W11} the migration and damping time scales are both at least an order of magnitude longer, so that their effects of type 1 
migration are very weak.\\

To compare our migration times with earlier results, we note that for an Earth-sized body at 1~AU our nominal disc parameters yield 
$t_{\rm wav} \sim 1700$~yr and $t_m \sim 180$~kyr, which are much longer than the values of \cite{tanaka02} and \cite{TW04} because 
our disc is initially hotter. The wave time scales as $(H/r)^4$ so that a factor of 1.5 in $H/r$ will result in a factor 5 
in the migration time scale, while $t_{\rm m} \propto (H/r)^2$ so the effect of the disc temperature is weaker. As the disc evolves 
the temperature decreases, speeding up type 1 migration, but so does the surface density, slowing it down. In general, the effect of 
type 1 weakens with time.\\

Thus, the effect of type 1 migration should be relatively weak in our disc model compared to the traditional results of 
\cite{tanaka02} and \cite{TW04}. Apart from the migrating force, the eccentricity and inclination damping forces from the disc will 
also cause some inward migration of the embryos due to angular momentum loss, and we expect them to migrate inwards of the order of 
0.1~AU over the lifetime of the disc. In Fig.~\ref{fig:torqueue} we plot a contour map of the normalised torque on the planetary 
embryos as a function of their distance to the Sun and their mass at time zero of the age of the disc. The normalised torque is 
$\Gamma_{\rm n} = \Gamma_{\rm tot}/\Gamma_0$ where $\Gamma_0 = (m_{\rm emb}/M_\odot)^2(r/H)^2(v/r)^2\Sigma$. The migration rate is 
then $\dot{r}=-2r\Gamma_{\rm tot}/L$, where $L$ is the orbital angular momentum \citep{tanaka02}. In all the disc models the 
definition of the torque is always inward.

\section{Initial conditions}
\label{sec:nm}
For this project we have run a large sample of numerical simulations of the Grand Tack scenario. These simulations are categorised 
into four large sets, with further subdivisions therein. \\

All simulations start with a high number of small planetesimals, planetary embryos, and the gas giants Jupiter and Saturn. We do not 
include Uranus and Neptune in any of the simulations, because they do not have any immediate effects on the formation of the 
terrestrial planets \citep{W11}. In all simulations we chose to not take account of the effect of Saturn's mass growth. According to 
\cite{W11} and \cite{JM14}, other effects, such as the radial evolution of Saturn and the gas giant migration time scale, did not 
substantially change the final terrestrial planet systems. Given the high number of free parameters, we decided to follow \cite{W11} 
to cases where Jupiter is assumed to have its current mass and is initially placed on a near-circular orbit at 3.5 AU. A fully-grown 
Saturn is placed in the 2:3 resonance with Jupiter at 4.5 AU. During the first 0.1 Myr, Jupiter and Saturn migrate from their initial 
locations (3.5 AU and 4.5 AU) to the tack locations (either 1.5 AU and 2.0 AU for Jupiter), respectively. For the next 5 Myr, Jupiter 
and Saturn migrate out to $\sim$ 5.4 AU and $\sim$ 7.5 AU, which are appropriate initial conditions for late giant planet migration 
models \citep{M07}. \cite{W11} demonstrated that the migration speed of the gas giants has almost no influence on the final 
architecture of the terrestrial system. We therefore used the same linear inward migration of the gas giants with a time-scale of 
0.1~Myr, followed by outward migration via an exponential prescription with an e-folding time of 0.5~Myr. These time scales are 
comparable to the typical rate of Type 2 migration \citep{LP86}.\\

To compare the effects of different formation models on the architecture of the terrestrial planets, we use various initial 
distributions of embryos and planetesimals.

\subsection{Equal mass embryo initial conditions}
For the first two large sets of simulations we employ the initial conditions of the embryos and planetesimals from \cite{JM14}, but we 
use our model for the protoplanetary disc. These simulations were run because we want to directly compare the results of our modified 
simulations -- employing a different protoplanetary disc, including type 1 migration and a realistic mass-radius relationship -- with 
theirs. All of the simulations in this set use equal mass embryos initially situated between 0.7~AU and 3~AU. The embryos were 
embedded in a disc of planetesimals. The surface density of embryos and planetesimals both scaled with heliocentric distance as 
$r^{-3/2}$. Following \cite{JM14} the total mass ratio of embryos and planetesimals in this inner disc is either 1:1, 4:1 or 8:1, with 
the individual embryo masses being either 0.025, 0.05 or 0.08~$M_\oplus$. The equal mass embryo assumption appears to agree with a 
pebble-accretion scenario of embryo formation \citep{Morby15}(Levison et al., 2015) rather than the more traditional oligarchic growth 
scenario \citep{ki98}, although which scenario is favoured is still under considerable debate. To mimic the coagulation evolution of 
the solids in the disc, we follow \cite{C06} and calculate the age of the disc is 0.1~Myr when embryos have a mass of 
0.025~$M_\oplus$, it is 0.5~Myr when the embryos have a mass of 0.05~$M_\oplus$ and 1~Myr when the embryos have a mass of 
0.08~$M_\oplus$.\\

Following \cite{JM14} again, the total mass in solids in the inner disc (embryos and planetesimals) is 4.3~$M_\oplus$~when the total 
mass ratio between embryos and planetesimals is 1:1, 5.3~$M_\oplus$~when the mass ratio is 4:1 and 6.0~$M_\oplus$~when the mass ratio 
is 8:1. \cite{JM14} further argued these different initial disc masses were necessary to keep the post-migration mass in solids 
between 0.7~AU and 1~AU close to 2~$M_\oplus$. Therefore, the surface density in solids between these different initial conditions 
increases with increasing total embryo to planetesimal mass ratio. The number of planetary embryos ranged from 29 to 213 depending on 
their initial mass and total mass ratio. We kept the number of planetesimals at 2000, regardless of their total mass. The initial 
densities of the planetesimals and embryos was 3~g~cm$^{-3}$ \citep{W11}. The permutations of these initial conditions results in nine 
individual sets of simulations.\\

Following \cite{M15}, for most of the simulations we also added an outer disc of planetesimals. This disc consists of 500 
planetesimals with a total mass of 0.06~$M_\oplus$. These planetesimals are to some degree considered responsible for volatile delivery 
on the otherwise dry terrestrial planets \cite{M15}. The outer planetesimals are distributed between 5~AU and 9~AU. The 
eccentricities and inclinations of both embryos and planetesimals are randomly chosen from a uniform distribution between 0 and 0.01 
and 0 to 0.5$^\circ$. The other angular orbital elements were chosen uniformly at random from 0 to 360$^\circ$. We ran two sets of 
these nine permutations, one with a tack at 1.5~AU as in \cite{W11}, and one with a tack at 2~AU as described in \cite{M15}.

\subsection{Oligarchic initial conditions}
Apart from simulating the formation of the terrestrial planets from a disc of equal-mass embryos and planetesimals we also run a 
second 
set of simulations where the initial conditions are reminiscent of the traditional oligarchic growth scenario \citep{ki98}. To set up 
our simulations, we used the semi-analytical oligarchic approach of \cite{C06}. In that work, the mass of embryos increases up to 
their isolation mass as
\begin{equation}
 m_{\rm p}(t)=m_{\rm iso}\tanh^3\Bigl(\frac{t}{\tau}\Bigr),
\end{equation}
where $m_{\rm iso}=2\pi a\Sigma_{\rm s} b$ is the isolation mass at semi-major axis $a$ for embryos spaced $b$~AU apart embedded in a 
disc with solid surface density $\Sigma_{\rm s}$. Here $\tau$ is the growth time scale which is a complex function of the semi-major 
axis, embryo spacing, solid surface density and radii of planetesimals that accrete onto the embryos. The growth time scale is given 
by \citep{C06}
\begin{equation}
 \tau = \frac{2e_{\rm Hi, eq}^2}{A}
\end{equation}
where
\begin{eqnarray}
 e_{\rm Hi, eq}^2 &=& 2.7^2\Bigl(\frac{r_{\rm c} \rho}{b C_D a \rho_{\rm gas}}\Bigr)^{2/5} \\
 &=& 17.5\Bigl(\frac{r_{\rm c}}{10\,{\rm km}}\Bigr)^{2/5}\Bigl(\frac{\rho}{2.5\,{\rm g~cm}^{-3}}\Bigr)^{2/5}
\Bigl(\frac{b}{10\,R_{H}}\Bigr)^{-2/5}
\Bigl(\frac{\rho_{\rm gas,0}}{1.4 \times 10^{-9}\,{\rm g~cm}^{-3}}\Bigr)^{-2/5}
\Bigl(\frac{a}{1\,{\rm AU}}\Bigr)^{2/5\alpha+1/5-1/5\beta} \nonumber,
\end{eqnarray}
assuming the drag coefficient $C_D=2$ and
\begin{eqnarray}
\frac{1}{A}&=&\frac{b^{1/2}P\rho^{1/3}M_\odot^{1/6}}{31.7\Sigma_{\rm s}^{1/2}} \\
 &=& 
30.1~{\rm kyr}\Bigl(\frac{b}{10\,R_{H}}\Bigr)^{1/2}\Bigl(\frac{\rho}{2.5\,\rm{g~cm}^{-3}}\Bigr)^{1/3}
\Bigl(\frac{\Sigma_s}{10\,\rm{g~cm}^{-2}}\Bigr)^{-1/2}\Bigl(\frac{a}{1\,\rm{AU}}\Bigr)^{3/2-1/2\alpha}M_\odot^{1/6}.
\end{eqnarray}
Here $\alpha$ is the slope of the solid surface density, $\beta$ is the slope of the temperature, $\rho_{\rm gas,0}$ is the gas 
density at 1~AU in the midplane, $r_{\rm c}$ is the radius of planetesimals, $\rho$ is their density and $P$ is the orbital period 
\citep{C06}. Here we used the fact that the gas density profile depends on the scale height and gas surface density which yields 
$\rho_{\rm g} \propto a^{-\alpha -3/2 +1/2\beta}$. Combining all the above gives
\begin{eqnarray}
\tau&=&527\,{\rm kyr}\,\Bigl(\frac{b}{10\,R_{H}}\Bigr)^{1/10}\Bigl(\frac{\Sigma_s}{10\,{\rm 
g\,cm}^{-2}}\Bigr)^{-1/2}\Bigl(\frac{\rho}{2.5\,{\rm g\,cm}^{-3}}\Bigr)^{11/15}\Bigl(\frac{r_{\rm c}}{10\,{\rm km}}\Bigr)^{2/5} 
\nonumber \\
&\times&
\Bigl(\frac{\rho_{\rm gas,0}}{1.4 \times 10^{-9}\,{\rm g~cm}^{-3}}\Bigr)^{-2/5}
\Bigl(\frac{a}{1\,{\rm AU}}\Bigr)^{17/10-1/10\alpha-1/5\beta}M_{\odot}^{1/6},
\end{eqnarray}
where $\Sigma_s$ is the solid disc surface density at 1~AU and we assumed that $\Sigma_{\rm s}$ has the same radial dependence as the 
gas surface density. This derived value of $\tau$ and the subsequent growth of any embryo near 1~AU agrees well with Figure~1 in 
\cite{C06}. Adopting $\alpha=3/2$ and $\beta=6/7$, the above equation is a reasonably steep function of semi-major axis at a fixed 
epoch because $\tau \propto a^{193/140} \sim a^{1.38}$. When we use the typically-assumed value $\beta=1/2$ then we have $\tau \propto 
a^{3/2}$. At the Earth's location the growth time for nominal parameters is $\sim$600~kyr while at Mars' current orbit the growth time 
is $\sim$1~Myr. This is somewhat shorter than that advocated by \cite{DP11} but is within error margins and is easily increased to 
1.8~Myr assuming accretion was caused by planetesimals of $\sim$50~km.\\

We constructed our initial disc of embryos and planetesimals as follows. First, we computed the total mass in solids between 0.7~AU 
and 3~AU assuming the surface density in solids is $\Sigma_s=7$~g~cm$^{-2}(a/1\,{\rm AU})^{-3/2}$. This setup is just the minimum mass 
solar nebula \citep{H81}. Second, following \cite{OI09}, we subsequently increased the solid density by a factor of 3 at the ice line, 
assumed to be static at 2.7~AU. Third, based on the results of \cite{ki98}, we imposed a spacing of 10 mutual Hill radii for the 
embryos, with the spacing computed assuming the embryos had their isolation masses. In other words, the semi-major axis of embryo $n$ 
is $a_n=a_{n-1}[1+b(2m_{\rm iso}/3M_\odot)^{1/3}]$ so that the embryo spacing nearly follows a geometric progression. Following 
\cite{C06} we assumed a planetesimal size of 10~km in computing the growth time scale of the embryos. Last, we entered the epoch 
at which Jupiter was assumed to have fully formed and began migrating. This was either 0.5 Myr, 1 Myr, 2 Myr or 3 Myr. Most embryos 
have only reached a fraction of their isolation mass and the remaining mass within their feeding annulus of 10 Hill radii taken up by 
planetesimals, with each planetesimal having a mass of 10$^{-3}$~$M_\oplus$. The eccentricities of the embryos and planetesimals 
followed a 
Rayleigh distribution with scale parameter equal to $(m_{\rm p}/3M_\odot)^{1/3}$. The inclinations also followed a Rayleigh 
distribution with a scale parameter equal to half of that of the eccentricities. The other angles were chosen uniformly at random 
between 0 and 360$^\circ$. All embryos and planetesimals had an initial density of 3~g~cm$^{-2}$. The most distant embryo was 
typically at 2.6~AU.

\section{Methodology}
\label{sec:methods}
\begin{figure}[t]
\resizebox{\hsize}{!}{\includegraphics[angle=-90]{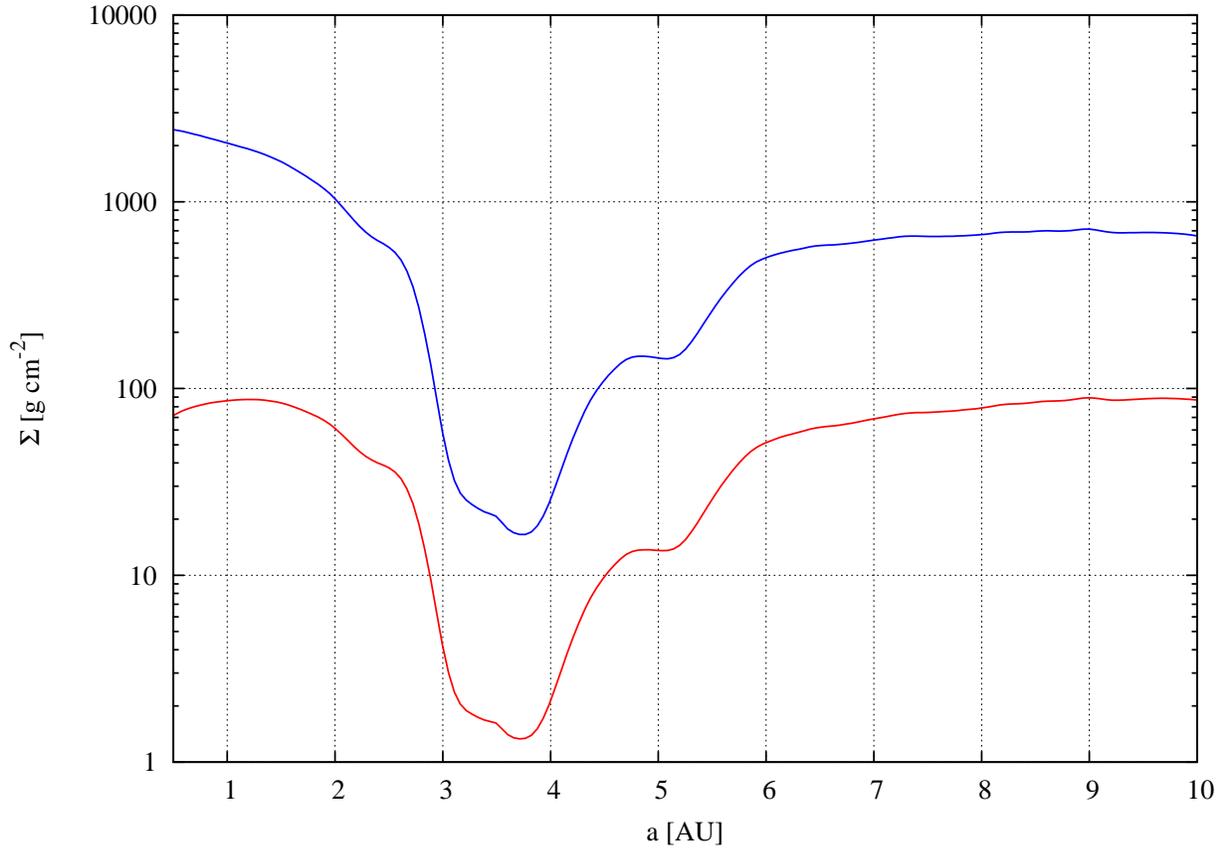}}
\caption{Density profile of the gas disc with Jupiter at 3.5~AU and the disc age is zero. The blue line is the disc we employed from 
\cite{B14} while the red line is the disc profile of \cite{W11}.}
\label{fig:gasdens}
\end{figure}

The gas giants, planetary embryos and planetesimals are simulated with the symplectic SyMBA integrator \citep{dll98} with a time 
step of 0.02~yr for 150~Myr. The end time of the simulations corresponds closely to the end of the purported Late Veneer (see next 
section). The migration of the gas giants was mimicked through fictitious forces described in \cite{W11}. We first employed the same 
gas profile as \cite{W11} with two large dips around the gas giants, which migrated inwards with these planets, but we modified the 
gas density profile during the computation so that it followed the $\Sigma(r) \propto r^{-1/2}$ surface density law of \cite{B14} 
rather than the Gaussian of \cite{W11}. The initial total mass of the disc was approximately 0.05~$M_\odot$. The initial disc profile 
is depicted in Fig.~\ref{fig:gasdens}. The blue line is the disc from \cite{B14}, whereas the red one is from \cite{W11}. Jupiter is 
assumed to be at 3.5~AU and the disc age is zero.
The planetary embryos experienced tidal damping of their eccentricities and inclinations and a negative torque from the gas disc as 
described in Section~2, and the orbits of the planetesimals evolve due to gas drag using the methods of \cite{B07}. 
Following \cite{W11} and \cite{JM14} for the purpose of the gas drag, we assumed each planetesimal had a radius of 50~km. These 
planetesimals are larger than the 10~km size assumed for oligarchic growth because we wanted to compare our simulations directly with 
\cite{JM14}. The gas drag routines of \cite{B07} were modified slightly to allow for a smooth transition between regimes when the 
Knudsen number crossed 1. We now have the following scheme:
\begin{itemize}
 \item When $\mathcal{M} > 2.727$, $C_D = 2$ for all $\mathcal{K}$ and $\mathcal{R}$;
 \item When $\mathcal{R} > 1000$, $C_D = 0.44+0.2098\mathcal{M}^2$ for $\mathcal{M} < 2.727$;
\item When $\mathcal{R} < 1000$
\begin{equation}
C_D = 0.2689\mathcal{M}^2 + 
\frac{24}{\mathcal{R}}\Bigl[1-\Bigl(\frac{\mathcal{M}}{2.727}\Bigr)^2\Bigr]\frac{1+0.15\mathcal{R}^{0.687}}{\sqrt{\mathcal{K}^2+1}}.  
\end{equation}
\end{itemize}
Here $\mathcal{M}$, $\mathcal{K}$ and $\mathcal{R}$ are the Mach, Knudsen and Reynold numbers of the planetesimals, and $C_D$ is the 
drag coefficient \citep{B07}. In addition to the drag coefficient routine, we made one further modification to the code.\\
SyMBA treats collisions between bodies as perfect mergers, preserving their density. This works well in most circumstances, but given 
that the mean density of Earth is 5.5 g~cm$^{-2}$ and not 3 g~cm$^{-2}$, we implemented the mass-radius relationship of \cite{S07} to 
make sure that the final planets have radii comparable to the current terrestrial planets so that their collisional cross sections are 
not artificially large. The relation we employed was
\begin{equation}
\log\Bigl(\frac{R_p}{3.3\,R_\oplus}\Bigr)=-0.209+\frac{1}{3}\log\Bigl(\frac{M_p}{5.5\,M_\oplus}\Bigr)-0.08\Bigl(\frac{M_p}{5.5\,
M_\oplus}\Bigr)^{0.4},
\end{equation}
where $R_p$ and $M_p$ are the radius and mass of the planetary embryo. This relation fits Mars, Venus and Earth well.\\
During the simulations we computed the mutual gravity between gas giants and embryos, but the planetesimals were not able to affect 
each other. This approximation was used to keep the CPU time within reasonable limits, and is justified because Jupiter clears the 
disc beyond 1~AU in 100~kyr. Planets and planetesimals were removed once they were farther than 100~AU from the Sun (whether bound or 
unbound) or when they collided with a planet or ventured closer than 0.2 AU from the Sun.\\

For each permutation of the equal mass embryo initial conditions of \cite{JM14} we ran 16 simulations (144 for each tack location), 
while for the oligarchic initial conditions we ran 16 simulations for each starting epoch (64 for each tack location). In total, we 
ran 416 simulations, categorised in Table~\ref{tab:sims}. The oligarchic simulations were run at the Centre for Computational 
Astrophysics at the National Astronomical Observatory of Japan, while the others were run at the Earth Life Science 
Institute at Tokyo Institute of Technology.
\begin{table}[t]
 \begin{tabular}{cccc}
 & Equal mass embryos & \\
 Embryo mass~[$M_\oplus$] & $M_{\rm emb}:M_{\rm pl}$ & Tack location & Migration epoch [Myr]\\ \hline \\
 0.025 & 1:1, 4:1 or 8:1 & 1.5~AU or 2~AU & 0.1 \\
 0.05 & 1:1, 4:1 or 8:1 & 1.5~AU or 2~AU & 0.5\\
 0.08 & 1:1, 4:1 or 8:1 & 1.5~AU or 2~AU & 1\\ \hline \\
& Oligarchic & \\
 & Migration epoch [Myr] & Tack location \\ \hline \\
 Oligarchic & 0.5 & 1.5~AU or 2~AU \\
 Oligarchic & 1 & 1.5~AU or 2~AU \\
 Oligarchic & 2 & 1.5~AU or 2~AU \\
 Oligarchic & 3 & 1.5~AU or 2~AU
 \end{tabular}
\caption{Summary of the individual sets of simulations. For each set of initial conditions we ran 16 simulations.}
\label{tab:sims}
\end{table}

\section{A measure of success}
\label{sec:success}
According to its founders, the Grand Tack model has booked several successes. These include, but are not limited to: the ability to 
reproduce the mass-orbit distribution of the terrestrial planets \citep{W11} (though mostly only for Venus, Earth and Mars), the 
compositional gradient and total mass of the asteroid belt \citep{W11}, the growth time scale of the terrestrial planets \citep{JW15} 
and in some cases the angular momentum deficit (AMD), spacing and orbital concentration of the terrestrial planets \citep{JM14}, 
and possibly the timing of the Moon-forming impact \citep{J14}. Some of these deserve further discussion before we outline our 
criteria for assigning success or failure to the individual simulations and the model itself.\\

\cite{J14} use simulations of terrestrial planet formation based on the Grand Tack model to place constraints on the time of the 
Moon-forming impact (a.k.a. Giant Impact or GI). They do this by requiring that, after the GI, the Earth subsequently 
accreted a further 0.5\% of its mass, which they claim is the best estimate compatible with the fraction of highly siderophile 
elements in the mantle used to account for the Late Veneer \citep{W09}. From their simulations they arrive at a timing of 
95$\pm$32~Myr. This is in agreement with the preferred Moon-forming time from hafnium-tungsten and samarium-neodymium geochronology, 
though on the higher end \citep{K05,T07}. It is also in agreement with recent dynamical modelling and $^{40-39}$Ar age compilations for 
the HED meteorites that likely originate from asteroid 4 Vesta \citep{B15}. In most of the simulations presented by \cite{JM14}, 
however, the last giant 
impact occurs much earlier than their preferred value. \cite{OB14} pointed out that such an early impact violates the constraint posed 
by the amount of highly siderophile elements in the Earth's mantle used to define the Late Veneer in the first place. That said, a 
late accretion of 1\% is still entirely within reason, and it may have been even higher: \cite{A13} argue that upwards of 4\% of 
Earth's mass was added to the planet by the Late Veneer. Their arguments are based on the timing and amount of water delivery, and the 
vaporisation of volatiles during accretion, throughout which a high fraction of impactor material is lost. Such a substantial amount 
of post-giant impact accretion would naturally push the epoch of the Moon-forming impact much further back in time, and leaves the 
timing issue once again wide open. A different approach or chronometer may be needed. \\

\cite{m12} argues for an Earth formation time shorter than 50~Myr. In a subsequent study \cite{AM14} use iodine, plutonium and xenon 
isotopic data to suggest that the closure time of the Earth's atmosphere is $40^{+20}_{-10}$~Myr, which would naturally coincide with 
the Moon-forming event. The formation time suggested by \cite{AM14} is in excellent agreement with the hafnium-tungsten dates of 
\cite{K05} (40$\pm$10~Myr), but on the lower end of that advocated by \cite{T07} (62$^{+92}_{-30}$~Myr) and \cite{H08} (70-100~Myr). 
In summary, the timing of the Moon-forming event is still a topic of ongoing debate. It appears that radiogenic dating results in an 
earlier GI time than suggested by the simulations of \cite{J14}, although the simulation results depend sensitively on the assumed 
amount of subsequent accretion. Most of the reported ages agree within error bars, but the range remains tens of millions of years.\\

Since individual simulations are chaotic and show a great variety in outcomes \citep{JM14}, we impose criteria the model must adhere 
to, which are listed below. In what follows, we define the Mercury, Venus, Earth and Mars analogues to have masses and semi-major axes 
within the ranges 
$(0.025\,M_\oplus<m<0.1\,M_\oplus,0.27\,\au<a<0.5\,\au)$,
$(0.4\,M_\oplus<m_p<1.2\,M_\oplus, 0.55\,\au<a<0.85\,\au)$, $(0.5\,M_\oplus<m_p<1.5\,M_\oplus,0.85\,\au<a<1.15\,\au)$ and 
$(0.05\,M_\oplus<m_p<0.15\,M_\oplus,1.3\,\au<a<1.7\,\au$). We require that objects in the region of the asteroid belt have their 
perihelia 
$q>1.6$~AU and their aphelia $Q<4.5$~AU. We want to add a note of caution. Since we employ initial conditions very similar to 
\cite{W11} and \cite{JM14} we expect to have a very low probability of reproducing the mass and semi-major axis of Mercury. One may 
then argue we should only base our analysis on the other three terrestrial planets, but we decided against doing so.\\

\cite{C01} introduced several quantities which describe the general dynamical properties of a planetary system. These are: 1) the AMD, 
given by

\begin{equation}
 {\rm AMD} = \frac{\sum_k \mu_k\sqrt{a_k}(1-\sqrt{1-e_k^2})\cos i_k}{\mu_k \sqrt{a_k}}
\end{equation}
where $\mu_k=m_k/M_\odot$. Second is the fraction of mass in the most massive planet ($S_m$). Third is a concentration parameter 
($S_c$), given by

\begin{equation}
S_c = {\rm max}\Bigl(\frac{\sum_k \mu_k}{\sum_k \mu_k [\log(a/a_k)]^2}\Bigr),
\end{equation}
and last, a mean spacing parameter ($S_H$), which is

\begin{equation}
 S_H = 2\sum_{k=1}^{N-1}\frac{a_{k+1}-a_{k}}{a_{k+1}+a_k}\Bigl(\frac{\mu_{k+1}+\mu_k}{3}\Bigr)^{-1/3}.
\end{equation}
Unlike \cite{C01} we use the mutual Hill sphere as the spacing unit. Ultimately we invoke the following criteria to determine success 
or failure of the Grand Tack model: statistically the resulting terrestrial systems must have their median AMD lower than the current 
value, a concentration parameter 107$\pm$67 (2$\sigma$), a mean spacing in Hill radii of 45$\pm$12 (2$\sigma$), a mass parameter 
$>0.5$, produce at least one Mars analogue (or a probability in excess of 5\% for a whole set) and the most massive planet must have a 
greater than 5\% probability of residing at 1~AU i.e. the cumulative semi-major axis distribution of the most massive planet must 
be lower than 0.95 at 1~AU. The ranges in the listed values of $S_c$ and $S_H$ are computed using a Monte Carlo method adopting 
the range of masses and semi-major axes of our terrestrial planet analogues stated above.

\section{Results: Tack at 1.5 AU}
\label{sec:results15}
In this section we present the results of our numerical simulations.

\subsection{Equal mass embryos}
We have run the same simulations as \cite{JM14} in order to compare our results directly with theirs, and to determine whether a 
different protoplanetary disc model, realistic mass-radius relationship and inclusion of type 1 migration will substantially change 
the properties of the resulting planets. Generally we find that most of our results are in good agreement. We shall not 
give a full comparison, but a systematic overview and highlight some similarities and differences.\\
\begin{figure}[t]
\resizebox{\hsize}{!}{\includegraphics[angle=-90]{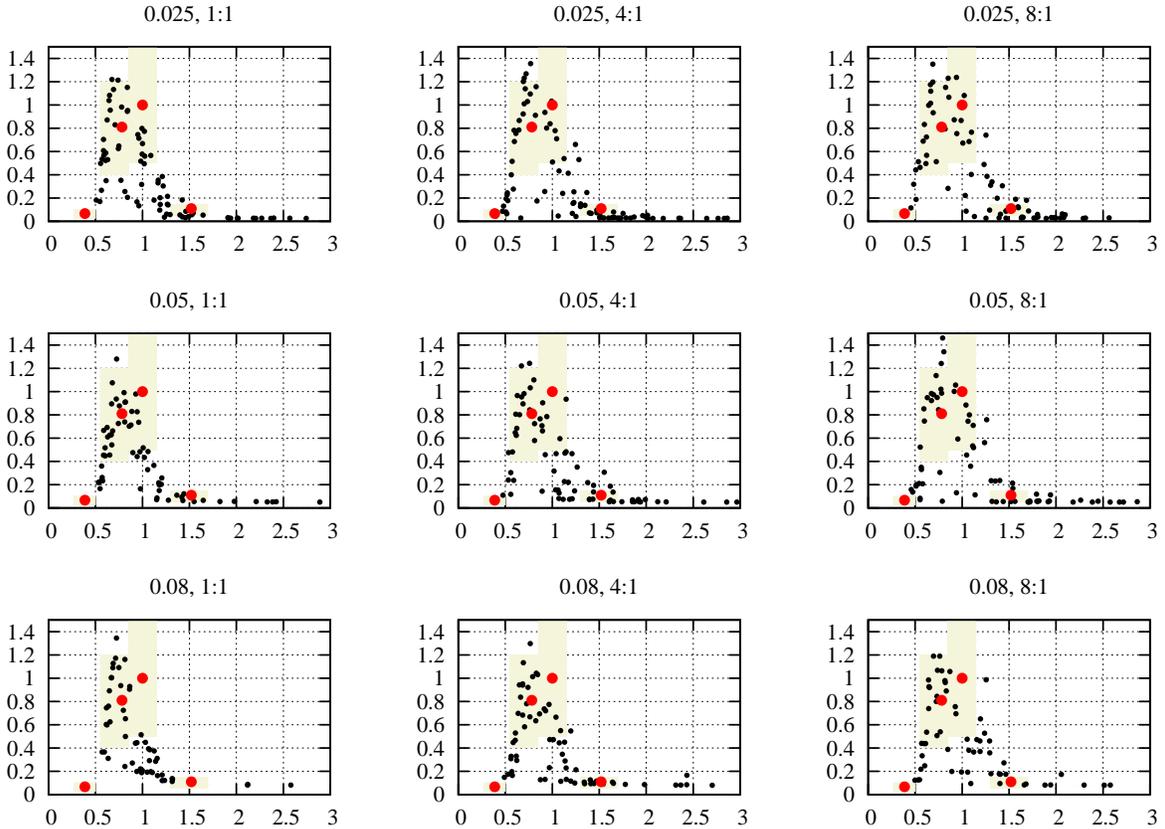}}
\caption{The final mass of the terrestrial planets [$M_\oplus$] versus their semi-major axis [AU]. The text above the panels indicates 
the 
embryo mass in Earth masses and the ratio of the total embryo to planetesimal mass. Beige regions indicate the range of our 
terrestrial planet analogues. Equal-mass embryo initial conditions with a tack at 1.5~AU.}
\label{fig:mvsajm15}
\end{figure}
\begin{figure}[t]
\resizebox{\hsize}{!}{\includegraphics[angle=-90]{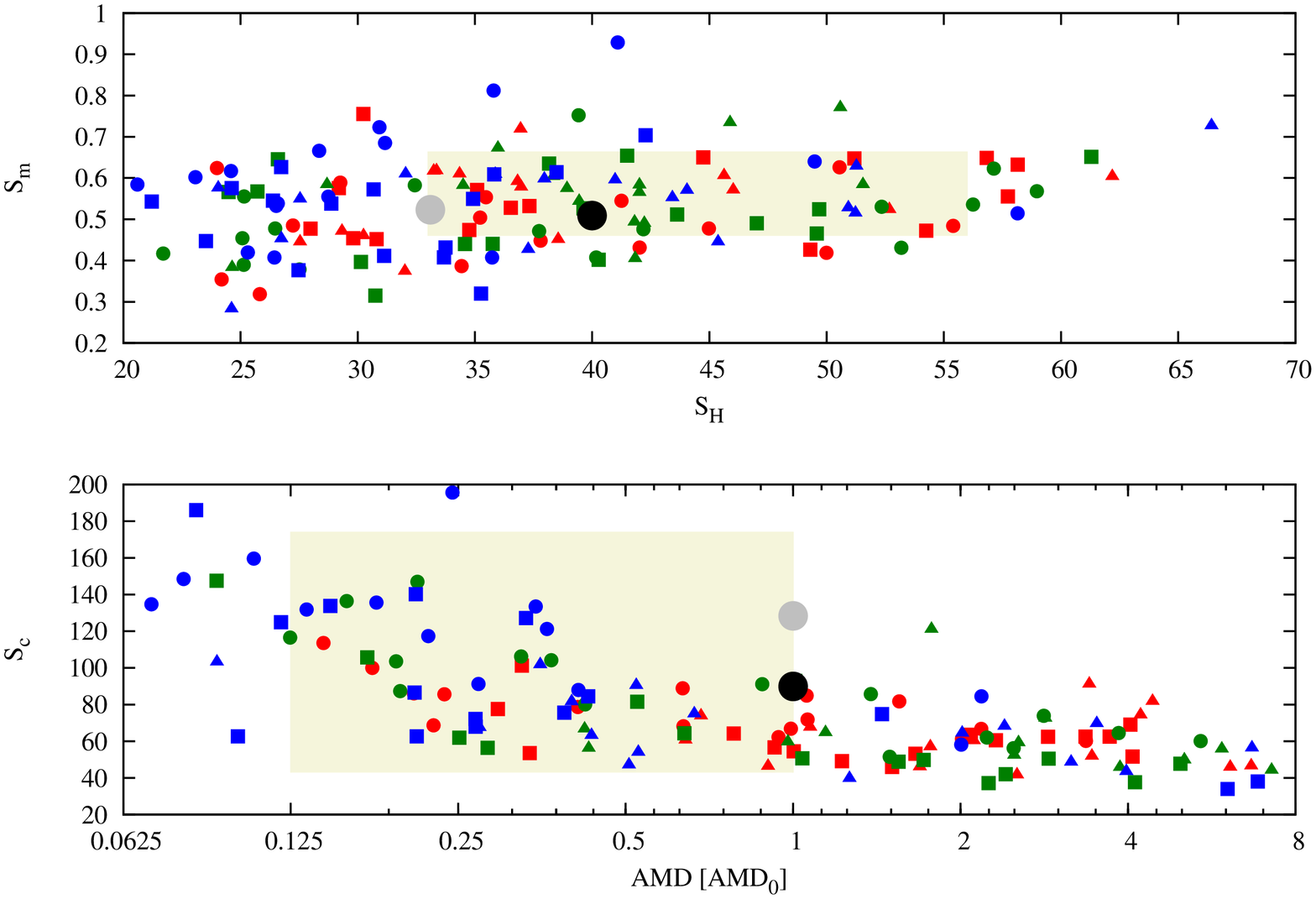}}
\caption{Top panel: Scatter plot of the spacing parameter $S_H$ vs the mass parameter $S_m$. Red symbols correspond to simulations 
with an initial embryo mass of 0.025~$M_\oplus$, green to initial embryo mass of 0.05~$M_\oplus$ and blue to initial embryo mass of 
0.08~$M_\oplus$. Bullets are for simulations with a total embryo to planetesimal mass ratio of 1:1, squares are for a 4:1 mass ratio 
and triangles for an 8:1 mass ratio. Bottom: Scatter plot of concentration parameter $S_c$ vs normalised AMD. Equal-mass embryo 
conditions with a tack at 1.5~AU.}
\label{fig:chambersjm15}
\end{figure}
In Fig.~\ref{fig:mvsajm15} we compare the mass of each terrestrial planet that formed in our simulation versus their semi-major axis. 
The actual terrestrial planets are depicted as red bullets. For the most part our results comport with Figure 1 in \cite{JM14}. 
In our simulations, however, the peak of the distribution is situated near Venus' current location while the region near Earth is 
empty in comparison. This is not the case for \cite{JM14} where the peak is in between these planets, encompasses both, 
and is probably caused by early type 1 migration of the embryos. The mean semi-major axis and mass of the most massive planet in 
our simulations are $\langle a_h \rangle = 0.769\pm0.109$~AU and $\langle m_h \rangle = 0.969\pm0.189$~$M_\oplus$, with almost no 
variation 
within error bars as a function of either embryo seed mass or total embryo to planetesimal mass ratio. Thus our Venus analogue is 
almost always more massive than the Earth analogue, and, with more than 95\% confidence, the position of the most massive planet is 
inconsistent with a location at 1~AU. This result is inconsistent with \cite{JM14} and we attribute this 
difference to our use of a distinctive model for the protoplanetary disc that has a a generally higher surface density, which causes 
stronger tidal damping, the inclusion of type 1 migration, and smaller planetary radii. Indeed, any material that is shepherded 
inwards by Jupiter will be at high eccentricity, which will be damped by interaction with the gas disc, which in turn will cause 
further inward migration since our gas disc has a higher surface density and a stronger damping than \cite{JM14}. We emphasise 
that the inward migration caused by the damping forces is generally stronger than the direct effect of the type 1 term, so that even 
using the non-isothermal prescription of \cite{P11} would not substantially change the outcome.\\

In summary, our setup causes a peak density in solids near Venus' current position rather than in between Earth and Venus as in 
\cite{JM14}. For this reason we ran another set of simulations with a tack location at 2~AU rather than the typical 1.5~AU to 
determine whether that would produce more Earth analogues. We also report a similar low success to \cite{JM14} in producing Mercury 
analogues which, like them, we attribute to the initial conditions.\\
\begin{table}[t]
 \begin{tabular}{ccccc}
 Embryo mass~[$M_\oplus$] & $\langle n \rangle$ & $\langle S_c \rangle$ & $\langle$ AMD $\rangle$ & $\langle S_H \rangle$ \\ 
\hline \\
 0.025 & 5.0 $\pm$ 1.1 & 66 $\pm$ 16 & 2.2 $\pm$ 2.2 (1.5) & 40 $\pm$ 11\\
 0.05 & 4.4 $\pm$ 1.3 & 71 $\pm$ 29 & 2.6 $\pm$ 3.3 (1.5) & 40 $\pm$ 11\\
 0.08 & 3.7 $\pm$ 1.0 & 97 $\pm$ 53 & 1.6 $\pm$ 3.1 (0.35) & 34 $\pm$ 10\\ \hline \\
 $M_{\rm emb}:M_{\rm pl}$ &  $\langle n \rangle$ & $\langle S_c \rangle$ & $\langle$ AMD $\rangle$ & $\langle S_H \rangle$ \\ \hline \\
 1:1 & 4.1 $\pm$ 1.1 & 101 $\pm$ 47 & 1.1 $\pm$ 1.5 (0.37) & 37 $\pm$ 13\\
 4:1 & 4.8 $\pm$ 1.2 & 72 $\pm$ 33 & 2.0 $\pm$ 3.0 (1.0) & 37 $\pm$ 10\\
 8:1 & 4.3 $\pm$ 1.4 & 62 $\pm$ 19 & 3.3 $\pm$ 3.4 (2.11) & 40 $\pm$ 10 
 \end{tabular}
\caption{Properties of the terrestrial systems with a tack at 1.5~AU and equal-mass embryos. We list the average number 
of planets, concentration parameter, AMD and average spacing with their standard deviations. Since the AMD distribution usually has a 
long tail, we list the median value in parentheses.}
\label{tab:nfin}
\end{table}
How do the resulting planets fare otherwise? In Fig.~\ref{fig:chambersjm15} we plot the spacing parameter $S_H$ versus mass 
parameter $S_m$ in the top panel and concentration parameter $S_c$ versus normalised AMD (normalised to the current value) in the 
bottom panel. The typical value of $S_c$ decreases with AMD because the higher eccentricities force the planets to be wider apart if 
they are to remain stable. In this figure and the ones that follow, red symbols correspond to simulations with an initial embryo mass 
of 0.025~$M_\oplus$, green to initial embryo mass of 0.05~$M_\oplus$ and blue to initial embryo mass of 0.08~$M_\oplus$. Bullets are 
for simulations with a 
total embryo to planetesimal mass ratio of 1:1, squares are for a 4:1 mass ratio and triangles for an 8:1 mass ratio. The large black 
bullet denotes the current terrestrial system while the grey bullet is for the system consisting of only Venus, Earth and Mars. The 
beige regions denote 2$\sigma$ regions around the mean values of $S_H$, $S_m$ and $S_c$. The scaled AMD range was chosen not to exceed 
1 because it will increase with time \citep{B13,L08}; the lower limit was chosen somewhat arbitrarily. Roughly 40\% of all simulations 
fall in either one of the beige regions, though only 10\% fall in both regions simultaneously. The results are summarised in 
Table~\ref{tab:nfin}. We reproduce the trend of \cite{JM14} that the AMD increases with a decrease in total planetesimal mass, though 
not with initial embryo mass. \cite{JM14} favour the cases with high embryo to planetesimal mass ratio, but the high resulting AMD is 
inconsistent with the dynamical evolution of the terrestrial planets \citep{B13,L08}. \cite{JM14} acknowledge the high AMD is a 
potential problem but suggest that fragmentation during embryo-embryo collisions could produce enough debris to damp the AMD through 
dynamical friction. It is not clear whether this can be sustained if collisional grinding is important. Further study is needed to 
support or deny this claim.\\

The terrestrial system consists of four planets. We find that the average final number of planets decreases with initial embryo mass 
and is mostly independent of the initial total mass ratio between the planets and embryos. The relatively high number of planets with 
low embryo seed mass is most likely skewed by stranded embryos in the asteroid belt or near Mars' current position. The concentration 
parameter $S_c$ increases with more massive embryos but decreases for a lower planetesimal mass, most likely because the AMD is higher 
and the planets need to be spaced farther apart to remain dynamically stable. \\

We define the average probability of producing a terrestrial planet analogue as the fraction of planets in the designated 
mass-semimajor axis bin divided by the total number of produced planets. For a Venus analogue this is $24\%\pm7\%$, for an Earth 
analogue it is $10\% \pm 3\%$ and for a Mars analogue it is $10\% \pm 4\%$. All of these values are above the 5\% threshold, but we 
see 
an over-abundance of Venus analogues consistent with the density pileup reported earlier. We produced a total of two Mercury analogues 
(out of 635 planets). \\

\begin{figure}[t]
\resizebox{\hsize}{!}{\includegraphics[angle=-90]{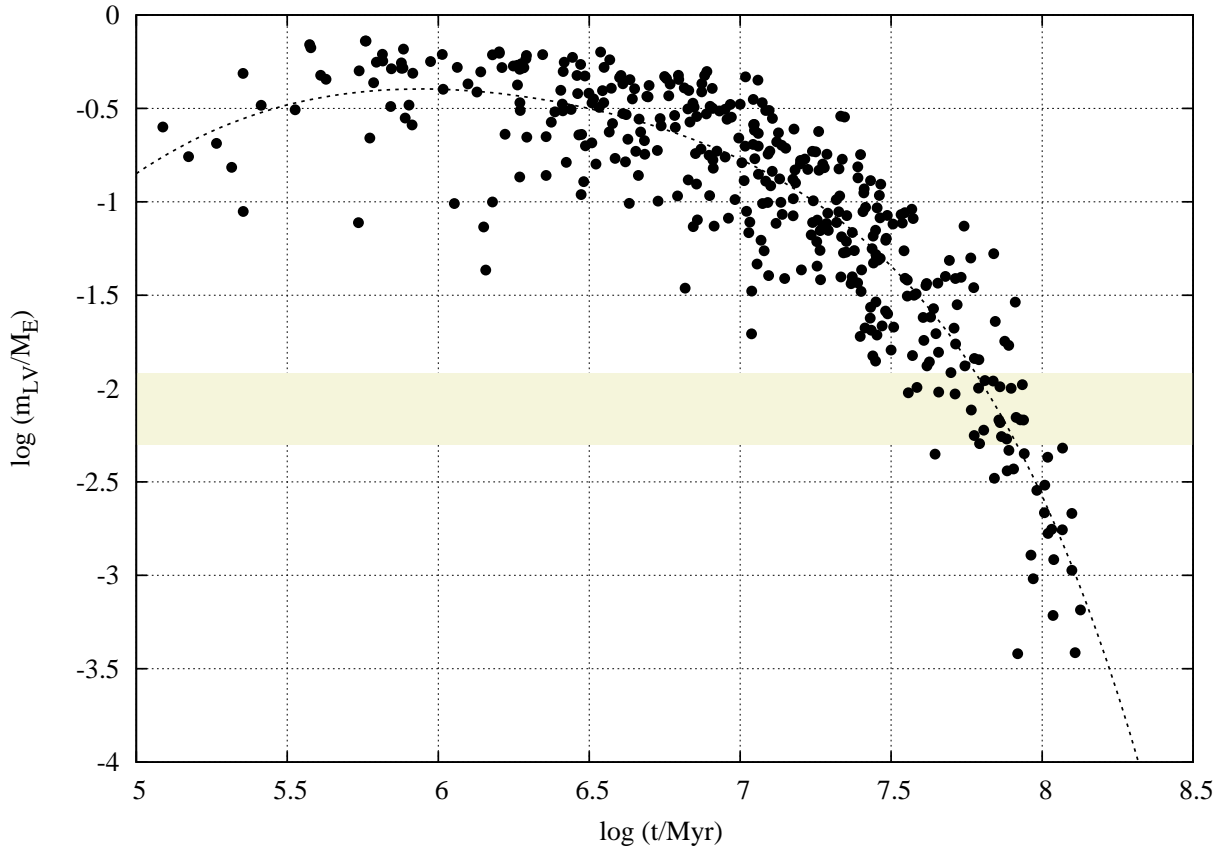}}
\caption{The last giant impact as a function of time versus subsequent accreted mass. The beige region comprises the amount 
constrained by highly siderophile elements in the Earth. Equal-mass embryo initial conditions with a tack at 1.5~AU.}
\label{fig:moonjm15}
\end{figure}
Thus far it appears that the only difference between our results and \cite{JM14} is the peak of the mass distribution being closer to 
the Sun than theirs, most likely because of our different disc model. Our results differ as well when we investigate the timing of the 
Moon-forming impact. Following \cite{J14} and \cite{JM14} again we compute the total amount of mass accreted by each planet between 
its last giant impact and the end of the simulation. We then plot this with a high-order polynomial best fit. The approximate timing 
of the Moon-forming impact is the intersection of the fit and some assumed Late Veneer mass \citep{J14}, which is at most a few 
percent of an Earth mass \citep{A13}. The lower the amount of assumed late accreted mass, the later the giant impact had to occur 
because less mass had to have been around to impact the Earth afterwards.\\

We plot our results in Fig.~\ref{fig:moonjm15}, and obtain a best-fit value of 64~Myr for the timing of the Moon-forming event 
assuming 1\% subsequent accretion. This is in good agreement with the Hf-W results from \cite{K05} and \cite{T07} but also 
sooner than what was suggested in the simulations of \cite{J14}. The difference in timing is most likely caused by us considering an 
accreted Late Veneer mass of 1\% rather than 0.5\%. However, the range of mass accreted after the giant impact is rather large. From 
the figure it is clear that this impact could have occurred anywhere between 30~Myr and 120~Myr, given the range of late 
accretion mass and uncertainties in the fit. Therefore we do not think that our simulations, or others such as \cite{J14} for that 
matter, can confidently predict the timing of the Moon-forming event with this method.\\

\begin{figure}[t]
\resizebox{\hsize}{!}{\includegraphics[angle=-90]{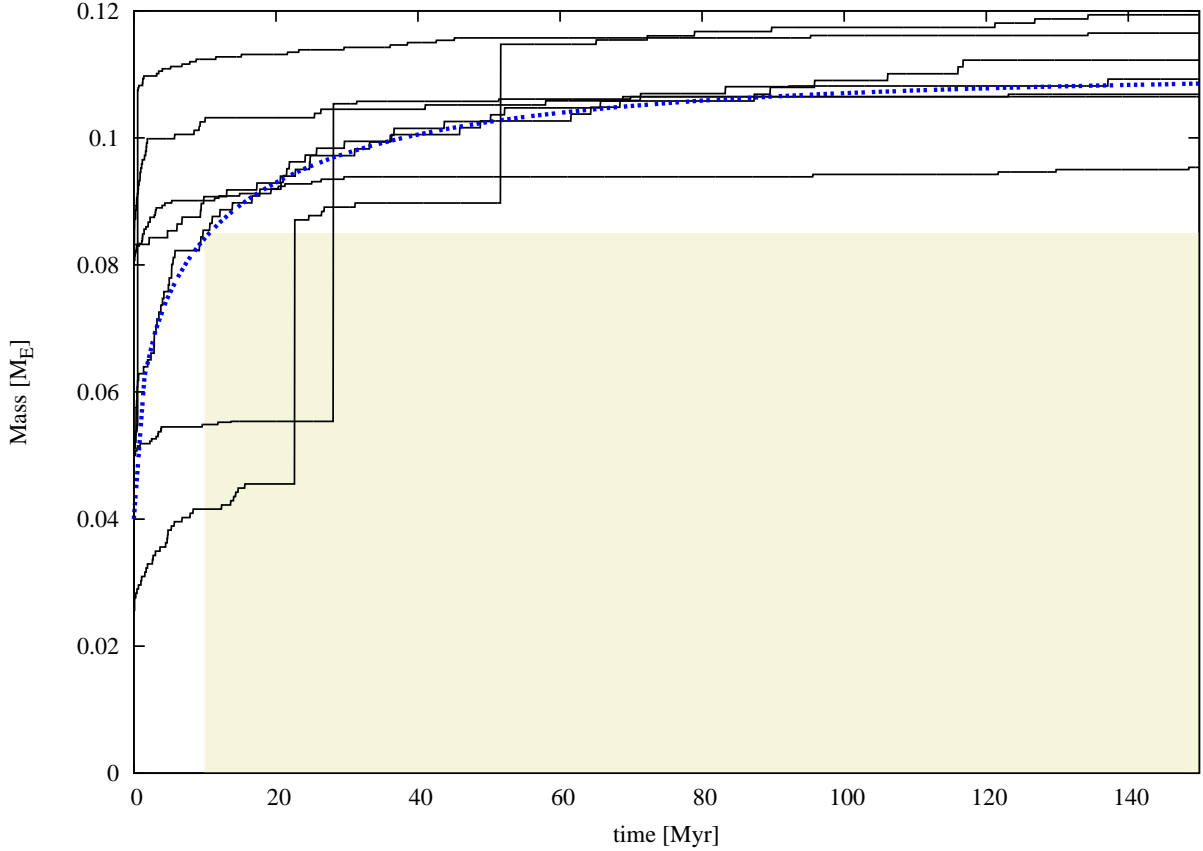}}
\caption{Evolution of mass with time for several Mars analogues. The beige region should be avoided because the formation 
time is inconsistent with the Hf-W chronometer \citep{NK07}. The blue curve is a Weibull cumulative distribution with e-folding time 
$\tau=10$~Myr, stretching parameter $\beta=0.5$ and embryo seed mass 0.04~$M_\oplus$.}
\label{fig:marsgrowth}
\end{figure}
Another issue that requires attention is the growth of Mars. \cite{JM14} conclude that it is very difficult to reproduce the 
rapid growth of Mars as advocated by \cite{DP11}. Figure~\ref{fig:marsgrowth} shows the evolution of the mass of several Mars 
analogues produced in our simulations as a function of time. The beige region should be avoided because the growth rate 
in this region is inconsistent with the Hf-W chronometer of Mars' formation \citep{NK07}. The blue dashed curve shows a stretched 
exponential growth function $m\propto m_{\rm Mars}(1-\exp[-(t/\tau)^\beta]$). We fit a seed mass of 0.04~$M_\oplus$, stretching 
parameter 
$\beta \sim 0.5$ and e-folding time $\tau \sim 10$~Myr. These values are nearly identical to the growth of Earth and Venus 
\citep{JW15}. Some Mars analogues experience early giant impacts with other embryos, substantially increasing their mass, but even 
then the final growth is slow and is inconsistent with the rapid growth advocated by \cite{DP11}, though still within limits of the 
Hf-W chronology of \cite{NK07}.\\

One last thing that has not been actively reported by either \cite{W11}, \cite{JM14} or \cite{OB14} is the amount of remaining mass in 
planetesimals. At the end of our simulations, we typically have a remnant mass in planetesimals of 
0.051\,$M_\oplus$$\pm$0.027\,$M_\oplus$, which is 
comparable to the total mass required to reproduce the Late Veneer \citep{R13}. The remnant mass depends on the original total mass 
ratio between planetary embryos and planetesimals. The simulations with an initial 1:1 ratio have a typical final mass of 
0.08\,$M_\oplus$ 
 while the 8:1 simulations typically have 0.01\,$M_\oplus$. The decay follows a stretched exponential with best fits
$\beta=0.43\pm 0.03$ and $\beta \log \tau = 0.40 \pm 0.04$, comparable to the results of \cite{JW15}. Thus, after 150~Myr of evolution 
the terrestrial planets would subsequently accrete an amount comparable to the Late Veneer. This could be problematic if the total 
accreted mass on the Earth after lunar formation is of the order of 1\% because the subsequent accretion would overshoot the accepted 
1\% value, but only by a small amount. 

\subsection{Oligarchic embryos}
\begin{figure}[t]
\resizebox{\hsize}{!}{\includegraphics[angle=-90]{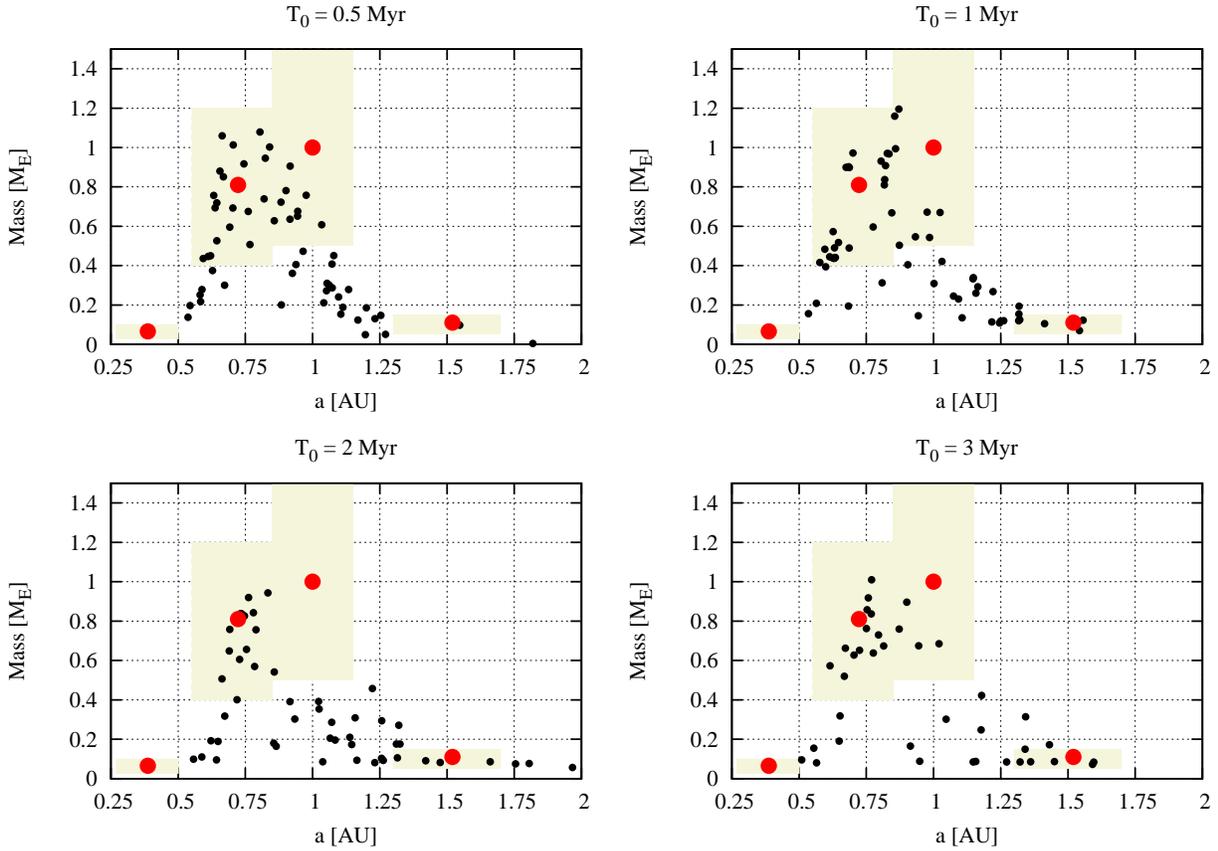}}
\caption{The final mass of the terrestrial planets versus their semi-major axis. The text above the panels indicates the time at 
which Jupiter and Saturn had formed and migrated into the inner solar system. Beige regions indicate the range of our terrestrial 
planet analogues. Oligarchic initial conditions with a tack at 1.5~AU.}
\label{fig:mvsao15}
\end{figure}
\begin{figure}[t]
\resizebox{\hsize}{!}{\includegraphics[angle=-90]{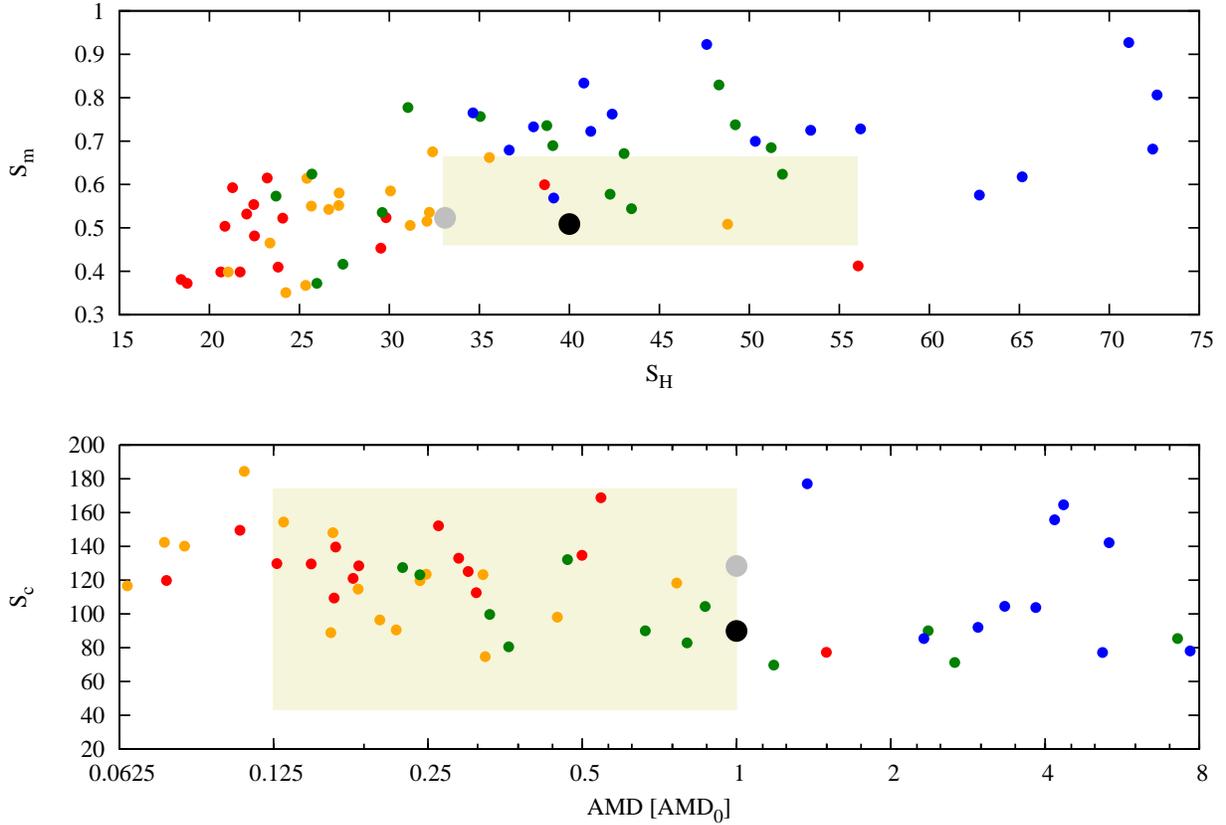}}
\caption{Top panel: Scatter plot of the spacing parameter $S_H$ vs the mass parameter $S_m$. Red symbols correspond to simulations 
with a disc age of 0.5~Myr, orange to initial disc age of 1~Myr, green dots have an initial disc age of 2~Myr and blue ones 3~Myr. 
Bottom: Scatter plot of concentration parameter $S_c$ vs normalised AMD. Oligarchic with tack at 1.5 AU.}
\label{fig:chamberso15}
\end{figure}

In this subsection we report the results of Grand Tack simulations with the oligarchic initial conditions. Since we do not expect 
substantial differences between this model and the equal mass embryo one of \cite{JM14}, we shall only report on the overall results. 
\\
\begin{table}[t]
 \begin{tabular}{ccccc}
 Disc age [Myr] & $\langle n \rangle$ & $\langle S_c \rangle$ & $\langle$ AMD $\rangle$ & $\langle S_H \rangle$ \\ 
\hline \\
 0.5 & 3.8 $\pm$ 0.7 & 133 $\pm$ 26 & 0.32 $\pm$ 0.33 (0.18) & 26 $\pm$ 9 \\ 
 1 & 3.7 $\pm$ 0.7 & 120 $\pm$ 28 & 0.23 $\pm$ 0.17 (0.18) & 29 $\pm$ 6 \\
 2 & 3.2 $\pm$ 0.6 & 127 $\pm$ 71 & 1.73 $\pm$ 2.64 (0.47) & 38 $\pm$ 9 \\
 3 & 2.6 $\pm$ 0.6 & 110 $\pm$ 64 & 7.1 $\pm$ 5.3 (4.3) & 52 $\pm$ 13\\ \hline
 \end{tabular}
\caption{Same as Table~\ref{tab:nfin} for the oligarchic initial conditions and a tack at 1.5~AU.}
\label{tab:nfino}
\end{table}
Figures~\ref{fig:mvsao15} and~\ref{fig:chamberso15} are the oligarchic equivalents of Figs.~\ref{fig:mvsajm15} 
and~\ref{fig:chambersjm15}. It appears that the correspondence with the real terrestrial system worsens as the time of the onset of 
migration increases. In the second figure the red dots are for the simulations where the disc age (time of the onset of 
migration) is 0.5~Myr. Orange dots are for a disc age of 1~Myr, green for 2~Myr and blue for 3~Myr. There are a few visible trends.
First, the final AMD value tends to increase with increasing disc age. This is unsurprising because the total mass in planetesimals 
decreases as the disc ages, so there is less mass to exert dynamical friction on the forming planets. Half of all systems are within 
the AMD-$S_c$ boundaries in the bottom panel, but only 11\% in the $S_H-S_m$ plot at the top, implying only 5\% fall into both regions 
simultaneously, lower than in the equal mass embryo case. The equal mass simulations have nearly uniform spacing anywhere from 20 to 
over 50 Hill radii. Systems with older disc ages are more widely spaced, while systems with younger disc ages -- and therefore lower 
embryo seed masses and more mass in planetesimals -- tend to be compact, with a typical spacing of 20 Hill radii, reminiscent of 
extrasolar systems \citep{FM13}. The older systems also tend to have fewer planets, and these planets all appear to be of similar, 
sub-Venus masses because we observe a trend of a decreasing number of planets with older disc ages. The mean number of 
planets as a function of disc age are listed in Table~\ref{tab:nfino}.\\

Visually the results from the oligarchic model appear to be different from the equal mass embryo setup, but statistically the models 
are nearly identical (see Table~\ref{tab:nfino}). We report no Mercury analogues, a probability of $32\% \pm 3\%$ for Venus analogues, 
$10\% \pm 5\%$ for Earth analogues and $9\% \pm 5\%$ for Mars analogues. The location and mass of the most massive planet is $\langle 
a_h \rangle = 0.782\au\pm 0.089\au$ and $\langle m_h \rangle = 0.805M_\oplus\pm0.160M_\oplus$, on the low side for both quantities. 
Once again the semi-major axis of the most massive planet is statistically inconsistent with the Earth's. In addition, unlike the equal 
mass embryo case, we did not change the disc mass from one set of of simulations to the next, which could account for the lower mass of 
the most massive planet.\\

We point out that we only tested the oligarchic initial conditions for disc mass with a a surface density of $\Sigma_0=7$~g~cm$^{-2}$ 
at 1~AU. It is possible that a higher initial surface density would lead to higher planetary masses at the end of the simulations. 
That said, it is unclear whether a higher surface density would increase the typical spacing between the planets, because of the weak 
dependence of the Hill radius on the mass.\\
When investigating the timing of the Moon-forming impact, we arrive at a time of 100~Myr, but once again the range is 
large, from 20~Myr to 120~Myr. The nominal value is a little later than favoured by the geochronology \citep{K05,T07} or the 
plutogenic-xenon arguments of \cite{AM14}. In any case, the mass left in planetesimals after 150~Myr of simulation is 
0.045\,$M_\oplus \pm 0.029\,M_\oplus$, comparable to the equal mass embryo case. Both the timing of the Moon-forming impact and 
remnant mass are 
statistically the same as for the equal mass embryo case. Lastly, the growth of Mars proceeds similarly to that depicted in 
Fig.~\ref{fig:marsgrowth}. In summary, both the oligarchic model and the equal mass embryo model are statistically identical within 
error margins, and we cannot favour one over the other. Further study is needed to distinguish the two.

\section{Results: Tack at 2 AU}
\label{sec:results20}
The simulations with a tack at 2~AU were performed to determine whether a more distant tack location would result in the Earth 
analogue being generally more massive than the Venus analogue and whether it improves the overall fit of the model with the current 
architecture of the terrestrial planets. Since much of the underlying dynamics are the same, we shall only report the highlights.

\subsection{Equal mass embryos}
\begin{figure}[t]
\resizebox{\hsize}{!}{\includegraphics[angle=-90]{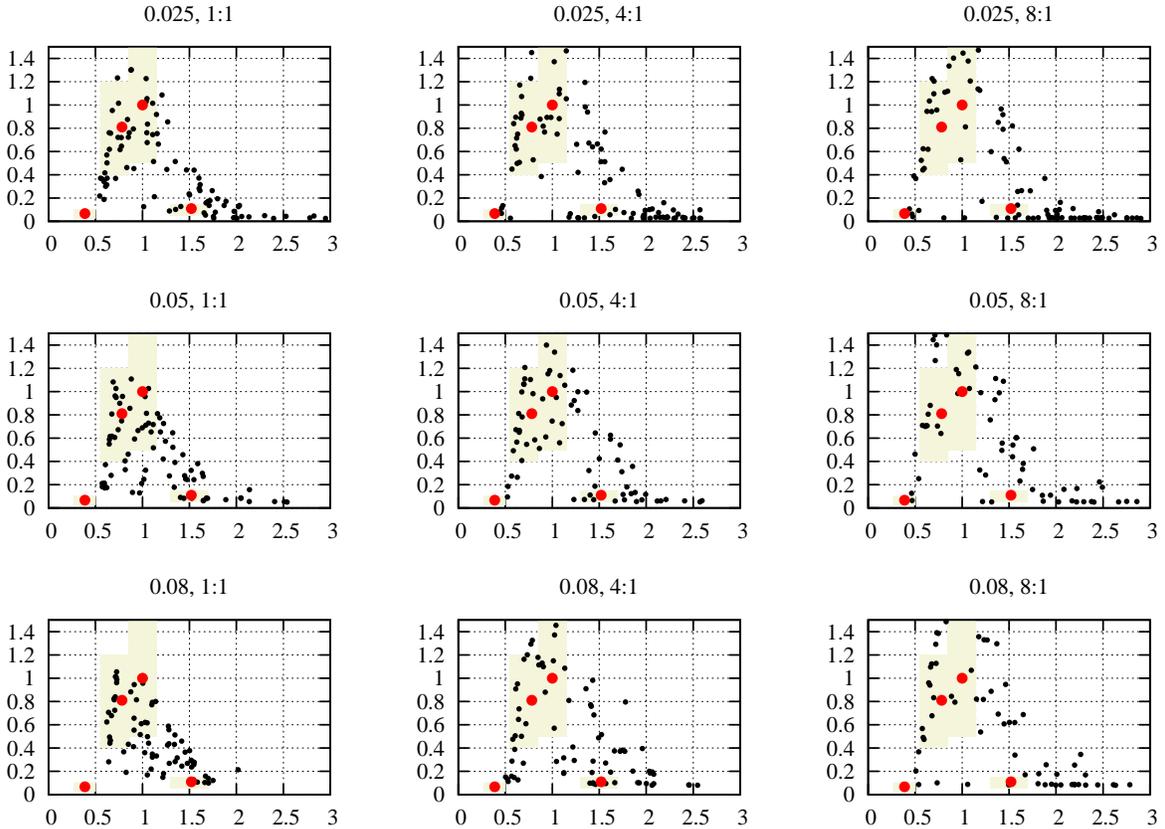}}
\caption{The final mass of the terrestrial planets versus their semi-major axis. The text above the panels indicates the embryo mass 
in Earth masses and the ratio of the total embryo to planetesimal mass. Beige regions indicate the range of our terrestrial 
planet analogues. Equal-mass embryo initial conditions with a tack at 2~AU.}
\label{fig:mvsajm20}
\end{figure}
\begin{figure}[t]
\resizebox{\hsize}{!}{\includegraphics[angle=-90]{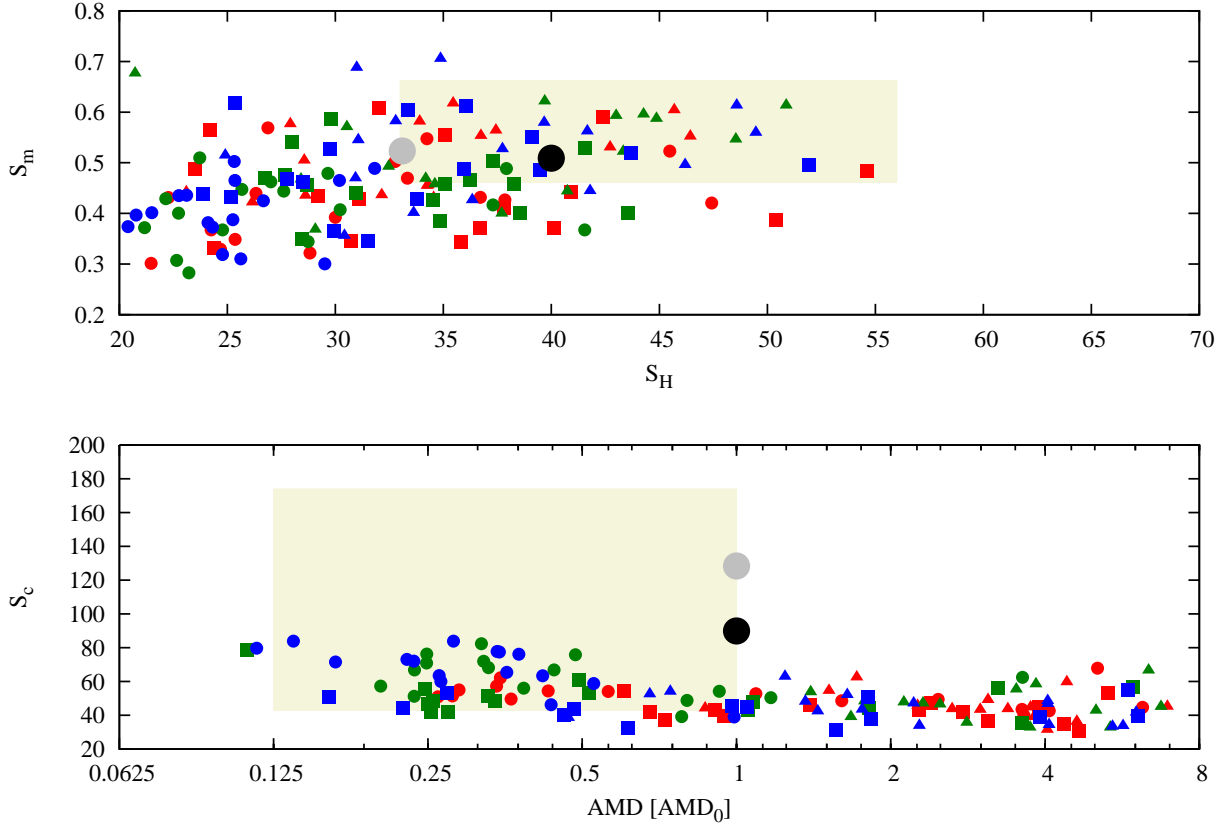}}
\caption{Top panel: Scatter plot of the spacing parameter $S_H$ vs the mass parameter $S_m$. Red symbols correspond to simulations 
with an initial embryo mass of 0.025$M_\oplus$, green to initial embryo mass of 0.05$M_\oplus$ and blue to initial embryo mass of 
0.08$M_\oplus$. Bullets 
are for simulations with a total embryo to planetesimal mass ratio of 1:1, squares are for a 4:1 mass ratio and triangles for an 8:1 
mass ratio. Bottom: Scatter plot of concentration parameter $S_c$ vs normalised AMD. Equal-mass embryo conditions with a tack at 
2~AU.}
\label{fig:chambersjm20}
\end{figure}
Figure~\ref{fig:mvsajm20} is a scatter plot of the final semi-major axis and masses of the planets produced in our simulations. The 
wider, more massive disc will naturally produce more massive planets over a wider range of heliocentric distances, and the plot
should be compared with Fig.~\ref{fig:mvsajm15}. There are two visual differences between the two sets of outcomes. First, the 
peak of the distribution is now farther out than with a tack at 1.5~AU. Indeed, the mean semi-major axis of the most massive planet is 
at $\langle a_h \rangle = 0.91\au \pm 0.19\au$, much closer to the current position of Earth than with a tack at 1.5~AU. The most 
massive planet now has a mean mass of $\langle m_h \rangle = 1.15M_\oplus \pm 0.26M_\oplus$, which is more massive than Earth but 
well within 
error margins. This increased mass is most likely caused by the accretion annulus being wider, having been truncated at 1.25~AU rather 
than at 1~AU. The outward tail is also caused by the same effect. We explore whether this also implies that a tack at 2~AU produces a 
better overall outcome.\\

First, we report that the average probability of producing a Venus analogue is $17\%\pm4\%$, an Earth analogue of $13\% \pm 5\%$ 
and a Mars analogue of $5\% \pm 3\%$. We also produce some Mercury analogues ($0.55\% \pm 0.83\%$). Overall, the production of Venus 
and Earth analogues is similar, while the production of Venus analogues was more than twice as high as Earth analogues when the tack 
occurred at 1.5~AU. The production of Mars analogues is low, at the threshold of acceptability, caused by the fact that we generally 
create more massive planets near Mars' current position than with a tack at 1.5~AU.\\

In comparison to the equal mass embryo simulations with a tack at 1.5~AU the planetary systems generated here have more planets on 
average. This is especially true for the cases with embryos of 0.025~$M_\oplus$ and a high mass in planetesimals. We list the number 
of 
planets and the standard deviation in Table~\ref{tab:nfin2}.
\begin{table}[t]
 \begin{tabular}{ccccc}
 Embryo mass~[$M_\oplus$] & $\langle n \rangle$ & $\langle S_c \rangle$ & $\langle$ AMD $\rangle$ & $\langle S_H \rangle$ \\ 
\hline \\
 0.025 & 5.4 $\pm$ 1.5 & 46 $\pm$ 8 & 3.2 $\pm$ 2.8 (2.6) & 33 $\pm$ 8\\
 0.05 & 4.5 $\pm$ 1.0 & 53 $\pm$ 13 & 2.4 $\pm$ 3.9 (0.93) & 33 $\pm$ 8\\
 0.08 & 4.4 $\pm$ 1.0 & 52 $\pm$ 15 & 2.1 $\pm$ 3.1 (0.74) & 32 $\pm$ 8\\ \hline \\
 $M_{\rm emb}:M_{\rm pl}$ &  $\langle n \rangle$ & $\langle S_c \rangle$ & $\langle$ AMD $\rangle$ & $\langle S_H \rangle$ \\ \hline \\
 1:1 & 5.0 $\pm$ 1.0 & 61 $\pm$ 13 & 1.0 $\pm$ 1.4 (0.36) & 28 $\pm$ 7\\
 4:1 & 4.9 $\pm$ 1.4 & 45 $\pm$ 9 & 2.6 $\pm$ 3.3 (1.1) & 34 $\pm$ 8\\
 8:1 & 4.5 $\pm$ 1.4 & 45 $\pm$ 9 & 4.1 $\pm$ 4.0 (3.1) & 36 $\pm$ 7 
 \end{tabular}
\caption{Same as Table~\ref{tab:nfin} for the equal mass embryos initial conditions and a tack at 2~AU.}
\label{tab:nfin2}
\end{table}
In Fig.~\ref{fig:chambersjm20} we once again plot the spacing parameter $S_H$ vs mass parameter $S_m$ in the top panel and 
concentration parameter $S_c$ vs normalised AMD in the bottom panel. A similar trend of decreasing $S_c$ with increasing AMD is 
visible, but not as pronounced. What is clear is that all of the systems have a concentration lower than or equal to that of the 
current terrestrial planets; none of them are higher (bottom panel). Since most systems have a spacing that is more compact than the 
current terrestrial planets (top panel), the lower concentration implies a lower variation in mass between the planets, which is 
generally what is observed in Fig.~\ref{fig:mvsajm20}. Things take a turn for the worse when we try to match the ranges of $S_c$, 
$S_m$, $S_H$ and the AMD simultaneously. We find that only 3/144 cases do so, which is much lower than the 5\% threshold we have 
adopted, and much lower than the 10\% reported earlier when the tack was at 1.5~AU. This low probability argues against a tack 
location at 2~AU being suitable to reproduce the current architecture of the terrestrial planets with the equal mass embryo initial 
conditions, despite the visual accuracy of fit. We generally find that $S_c$ is low while $S_H$ and the AMD are comparable to the 
case with a tack at 1.5~AU, suggesting that the mass distribution is narrower and the mass variations between planets are less 
extreme. Most of the concentration values are low but within the acceptable range.\\

The last two issues are the timing of the Moon-forming impact, which is at 60~Myr but with the same range (30~Myr to 120~Myr) as 
reported earlier, and a leftover planetesimal mass of $0.053M_\oplus \pm 0.04M_\oplus$, once again on the high end but in agreement 
with simulations having a tack at 1.5~AU. The simulations with an initial 1:1 ratio have a typical final mass of 0.1\,$M_\oplus$ while 
the 8:1 simulations typically have 0.02\,$M_\oplus$. The decay follows a stretched exponential with best fits $\beta=0.44\pm 0.03$ and 
$\beta \log \tau = 0.52 \pm 0.05$, which results in a slower decay than with a tack at 1.5~AU and explains the higher leftover mass.

\subsection{Oligarchic embryos}
\begin{figure}[t]
\resizebox{\hsize}{!}{\includegraphics[angle=-90]{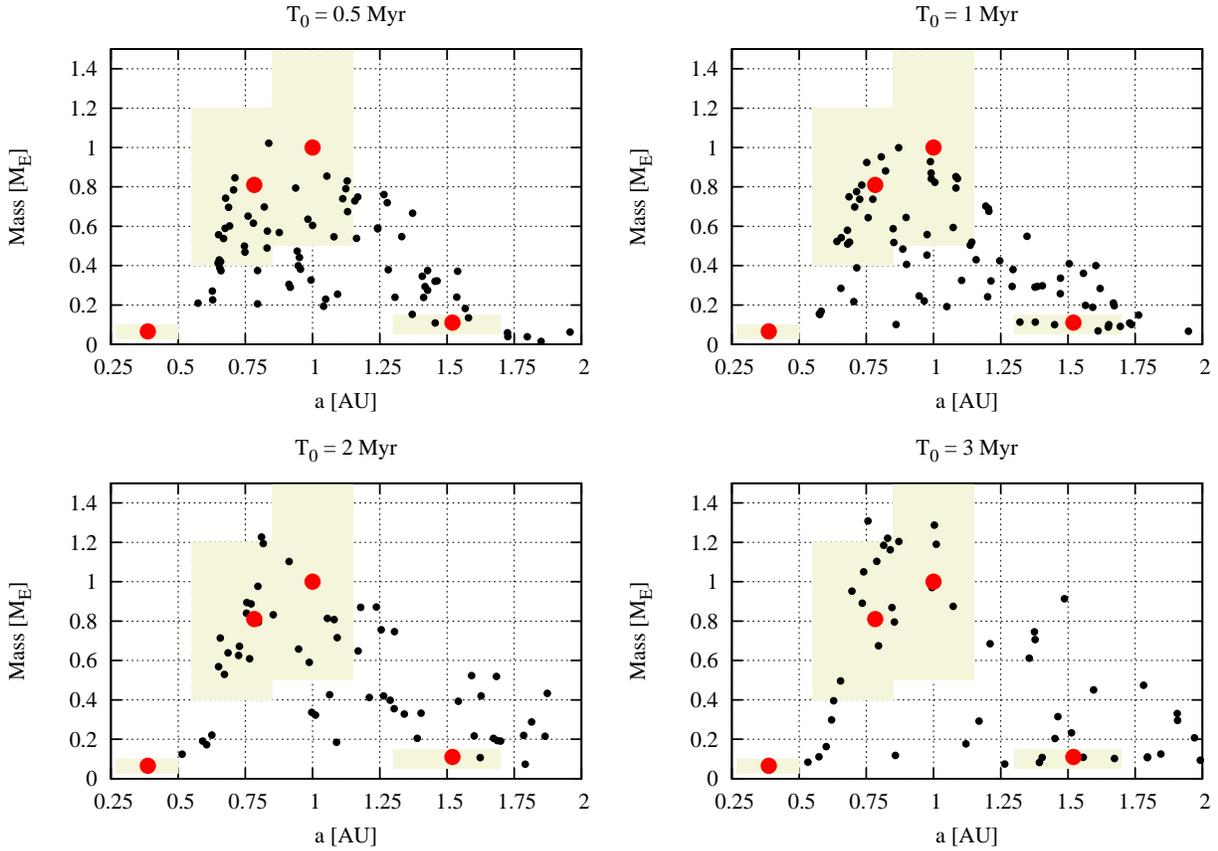}}
\caption{The final mass of the terrestrial planets versus their semi-major axis. The text above the panels indicates the time at 
which Jupiter and Saturn had formed and migrated into the inner solar system. Beige regions indicate the range of our terrestrial 
planet analogues. Oligarchic initial conditions with a tack at 2~AU.}
\label{fig:mvsao20}
\end{figure}
\begin{figure}[t]
\resizebox{\hsize}{!}{\includegraphics[angle=-90]{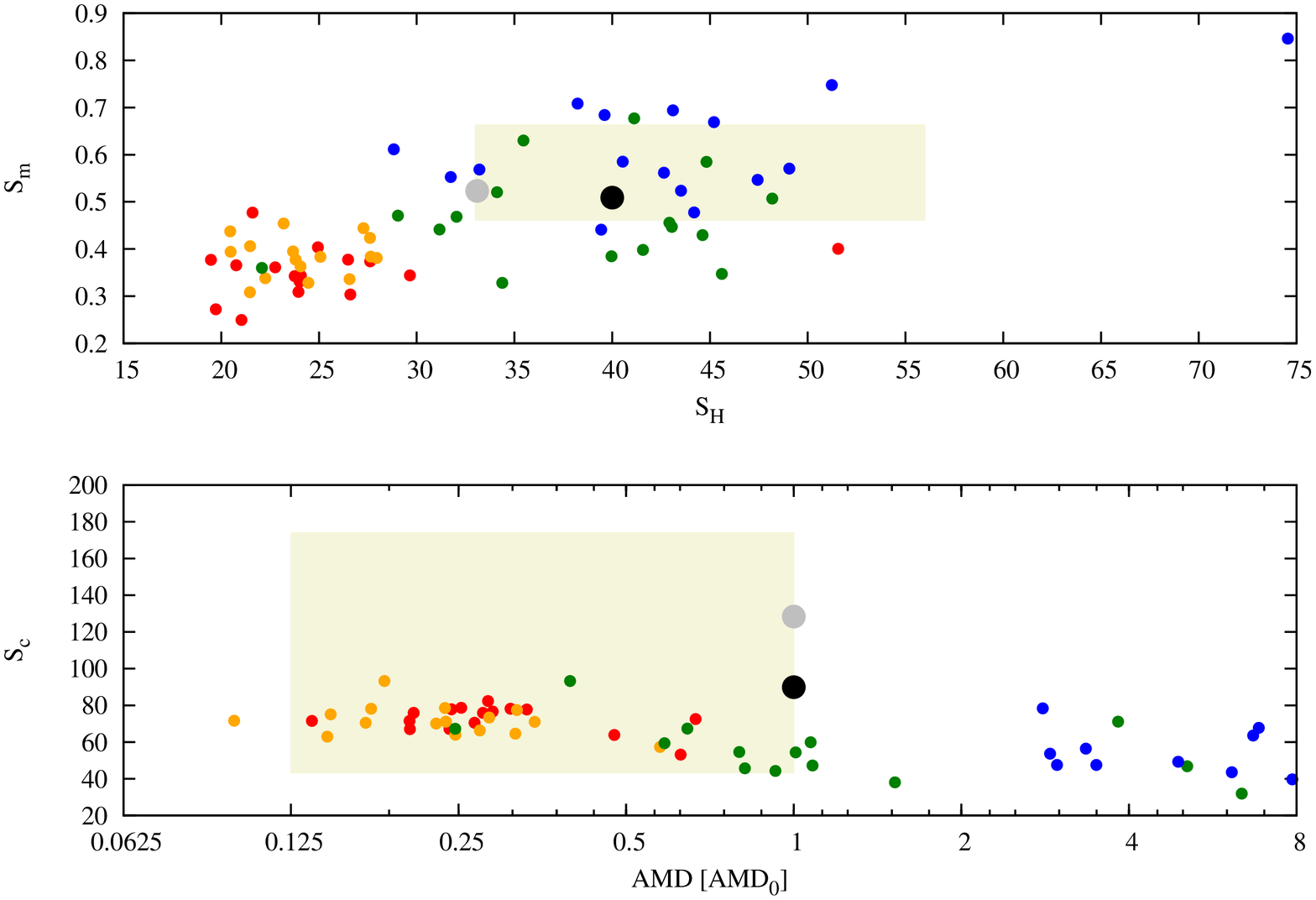}}
\caption{Top panel: Scatter plot of the spacing parameter $S_H$ vs the mass parameter $S_m$. Red symbols correspond to simulations 
with a disc age of 0.5~Myr, orange to initial disc age of 1~Myr, green dots have an initial disc age of 2~Myr and blue ones 3~Myr. 
Bottom: Scatter plot of concentration parameter $S_c$ vs normalised AMD. Oligarchic with tack at 2 AU.}
\label{fig:chamberso20}
\end{figure}
In the previous subsection we investigated whether or not a tack location at 2~AU would yield a better outcome for the overall 
architecture of the terrestrial planets than a tack at 1.5~AU in the case of equal mass embryo initial conditions. We concluded that 
it appears to be difficult for this combination of tack location and initial conditions to simultaneously reproduce the combined 
spacing, concentration, mass distribution and AMD of the terrestrial planets, despite generating more Earth analogues and the heaviest 
planet being closer to Earth's current location. It is now worth investigating whether the oligarchic system fares any better.\\

Figures~\ref{fig:mvsao20} and~\ref{fig:chamberso20} depict the relation between mass and semi-major axis, and spacing, 
concentration, mass and AMD distributions as usual. Once again we see a broader semi-major axis-mass distribution and an overall 
closer spacing and lower concentration. The concentration is, however, generally a little higher than in the equal mass embryo case. 
Indeed, we find that 10\% of the outcomes fall within both beige regions simultaneously, higher than for the equal mass embryo 
case and comparable to the simulations with a tack at 1.5~AU. The production of terrestrial planet analogues is similar to the equal 
mass embryo case above so that we tend to produce more planets on average than with a tack at 1.5~AU, but somewhat fewer than the 
equal 
mass embryo case with a tack at 2~AU. The final results are listed in Table~\ref{tab:nfino2}.\\

Similarly to the equal mass embryo case with a tack at 2~AU we produce Venus analogues $20\%\pm 3\%$ of the time, Earth analogues 
$13\% \pm 3\%$ of the time and Mars analogues with a probability of $6\% \pm 3\%$. The wider mass annulus also results in the heaviest 
planet having a mean semi-major axis of $\langle a_h \rangle = 0.96\au \pm 0.18\au$, but its mean mass remains a little low at 
$\langle m_h \rangle = 0.88M_\oplus \pm 0.18M_\oplus$, probably because we did not enhance the disc mass per set of simulations as 
we did for 
the equal mass embryo case. Once again the concentration value is low and we see $S_H \sim 25$ Hill radii for early disc ages and an 
increase in the AMD with disc age.\\

The Moon-forming impact occurs at 90~Myr, ranging from 30~Myr to 120~Myr depending on the amount of late accretion. The mass in 
leftover planetesimals is comparable to earlier simulations at $0.054M_\oplus \pm 0.038M_\oplus$.
\begin{table}[t]
 \begin{tabular}{ccccc}
 Disc age [Myr] & $\langle n \rangle$ & $\langle S_c \rangle$ & $\langle$ AMD $\rangle$ & $\langle S_H \rangle$ \\ 
\hline \\
 0.5 & 4.8 $\pm$ 0.8 & 73 $\pm$ 7 & 0.31 $\pm$ 0.14 (0.27) & 25 $\pm$ 7 \\ 
 1 & 4.9 $\pm$ 0.6 & 72 $\pm$ 8 & 0.24 $\pm$ 0.11 (0.24) & 24 $\pm$ 3 \\
 2 & 3.9 $\pm$ 0.8 & 53 $\pm$ 16 & 2.9 $\pm$ 3.6 (1.0) & 38 $\pm$ 7 \\
 3 & 3.6 $\pm$ 0.7 & 48 $\pm$ 13 & 9.7 $\pm$ 8.8 (6.7) & 43 $\pm$ 10\\ \hline
 \end{tabular}
\caption{Same as Table~\ref{tab:nfin} for the oligarchic initial conditions and a tack at 2~AU.}
\label{tab:nfino2}
\end{table}

\subsection{Summary}
The summary of our results is displayed in Table~\ref{tab:sum}. The criteria we consider important are whether the model can produce 
Mars with a probability higher than 5\%, whether the architecture of the system in terms of spacing, concentration, mass and AMD is 
consistent with the current planets, and whether the mass and semi-major axis of the heaviest planet are consistent with those of 
Earth. The Moon-forming impact is a less stringent criterion because of the inherent uncertainty in the late-accreted mass and the 
wide range of ages that appear as a result. All that matters is that it is within the error bars of the reported hafnium-tungsten 
ages, which it is, but it does not provide any further constraints on the model.\\

It appears that most initial conditions work to some extent. A tack at 2 AU is able to place the most massive planet near Earth's 
current location and thus being the only model to satisfy the semi-major axis constraint of the heaviest planet. This statistically 
rules out a tack at 1.5~AU with the constraints that we impose and the disc parameters and migration prescription that we used. 
However, the equal mass embryo case with a tack at 2~AU is problematic because of its low ability to reproduce the spacing, 
concentration, mass and AMD ranges simultaneously. We are confident this low probability is inherent in the model and not caused by 
sampling and therefore we also rule it out. This leaves us with the oligarchic model with a tack at 2~AU. The remnant mass in 
planetesimals is an obvious concern, but the somewhat low mass of the most massive planet when the tack occurred at 2~AU cannot be 
statistically rejected. The disc age in the oligarchic model that best matches all constraints is 2~Myr, which is a typical time 
range to form the gas giants.\\
\begin{table*}
 \begin{tabular}{cccccccc}
  Type & Tack [AU] & Mars & Architecture & $\langle a_h\rangle$ & $\langle m_h\rangle$ & $t_{\rm Moon}$ 
& Mass left [$M_\oplus$]\\ \hline \\
  Equal mass & 1.5~AU &  \cmark & \cmark & \xmark & \cmark & \cmark & 0.05$\pm$0.03 \\
  Equal mass & 2~AU &  \cmark & \xmark & \cmark & \cmark & \cmark & 0.05$\pm$0.04\\
  Oligarchic & 1.5~AU &  \cmark & \cmark & \xmark & \cmark & \cmark & 0.05$\pm$0.03\\
  Oligarchic & 2~AU & \cmark & \cmark & \cmark & \cmark & \cmark & 0.05$\pm$0.04
 \end{tabular}
 \caption{Summary of the results of the different sets of simulations. It is clear that a tack location of 1.5~AU has 
difficulty reproducing the current Solar System; a tack at 2~AU is preferred.}
  \label{tab:sum}
\end{table*}
We point out that the above conclusions, drawn from Table~\ref{tab:sum} are valid for the disc model that we have employed. We 
return to its implications in the next section. One thing we have not mentioned thus far is the total mass of material we emplace in 
the asteroid belt. This matter is discussed in detail in the next section.

\section{Discussion}
\label{sec:dis}
In this study we ran a high number of terrestrial planet simulations in the framework of the Grand Tack model with a range of initial 
conditions and two different tack locations. In the previous subsection we concluded that, despite using a different model for the 
protoplanetary disc, the inclusion of type 1 migration and a realistic mass-radius relationship, the outcomes of our simulations are 
broadly similar, though each setup has its own unique pros and cons.\\

We have decided to use a different disc model than the traditional setup of \cite{W11} and we have outlined our reasons for doing 
so. We find that a tack at 1.5~AU leaves too much mass near Venus' location. We attribute this to a combination of type 1 migration 
and inward shepherding by Jupiter, which decreases the semi-major axis of material down to below 1~AU but increases the eccentricity, 
so that further inward migration will ensue. This raises the question as to how sensitive our results are to the choice of disc 
model. This can only be answered by running more simulations with varying disc parameters, which is beyond the scope of the current 
study. We have used a relatively hot and puffy disc, with a scale height that is higher than traditional values \citep{H81,tanaka02, 
TW04} caused by viscous heating in the inner disc. This higher scale height causes slower embryo migration. Thus, making use of a 
colder, thinner disc would likely have exacerbated the overproduction of Venus analogues with a tack at 1.5~AU. One way to mitigate 
this problem is to use a much lower surface density, as was done by \cite{W11} and \cite{JM14} but their values seem artificial. A 
much lower surface density would decrease the migration rate of Jupiter and Saturn, and even though their migration rate does not 
appear to affect the final orbital architecture of the terrestrial system \citep{W11}, it does increase the difficulty for these 
planets to reach their final positions beyond 5~AU \citep{AM12}.  Thus, even when using a colder, less massive disc, we still 
expect to see an overproduction of Venus analogues when the tack occurred near 1.5~AU.\\

A second topic concerns the timing of the Moon-forming impact. We generally find agreement between our typical time of 60-90~Myr 
and geochronology. However, our uncertainties are typically 30~Myr or longer, suggesting the GI occurred anywhere from 30~Myr to 
120~Myr, which is what we have claimed in the previous sections. It is debatable whether this range implies anything meaningful. It is 
consistent with the value 95$\pm$32~Myr reported in \cite{J14}, even though they ran their simulations for a little longer and used a 
much lower amount of late accreted mass. In summary, we do not think that our simulations, nor those of \cite{J14}, can say anything 
meaningful about the timing of the GI beyond what is known from geochronology.\\

Another issue that requires discussion is the mass left over in planetesimals after planet formation. This is typically 0.05~$M_\oplus$ 
but can be as high as 0.1~$M_\oplus$. This leftover mass has implications for the cratering rates on Noachian Mars and the 
Pre-Nectarian Moon. The most efficient way to eliminate this material is through collision with the terrestrial planets. Ejection by 
the giant planets or collisions with the Sun is much more difficult. Preliminary simulations of this population of planetesimals 
indicate it decays slowly, following a stretched exponential with stretching parameter $\beta \sim 0.83$ and e-folding time $\tau \sim 
85$~Myr. A slow decay is preferred by lunar cratering records \citep{W14}, but a slower decay is necessary so as not to have late 
melting of the crust of the planets \citep{AB13}; Abramov \& Mojzsis (2016).\\

One solution may be for these planetesimals to grind themselves to dust and subsequently be lost through Poynting-Robertson drag or 
radiation pressure. The difficulty with this idea is that the high ratio of highly siderophile elements in the Earth and Moon suggest 
that the Late Veneer impactors were large (around 2000~km) \citep{B10}. If these impactors were large there is no reason to believe 
the impactors after the Late Veneer were substantially smaller. A simple argument is that Ceres is the only 1000~km body in the 
asteroid belt, and with a typical implantation probability of 0.1\% \citep{W11}, there should have been at least 1000 Ceres-sized 
bodies. With of the order of 5\% of the total mass remaining after 150~Myr we have 50 Ceres-sized bodies still present, with perhaps 
ten bodies the size of 4000~km. Since it was probable that the size distribution of the remnant planetesimals was shallow \citep{B10}, 
most of the mass is in the large bodies, and thus we consider it very unlikely that this mass was ground down by collisional erosion.\\

A quick estimate of the collisional time scale can be made with an $n\sigma v t=1$ argument, where $n$ is the number density of 
planetesimals, $v$ is their typical encounter velocity, and $\sigma=\pi r^2$ is their collisional cross section. The number density 
$n=M/(2 \pi^2 a^2 m\Delta a\sin i)$, where $M$ is the total mass of remnant planetesimals, $\Delta a$ is the width of the annulus 
in which the planetesimals are situated and $m$ is their individual mass. We then have
\begin{equation}
 \frac{3Mv t}{8\pi^2 a^2 \Delta a \sin i \rho r}=1.
\end{equation}
From our simulations we find that the typical semi-major axis of these planetesimals is 1.5~AU and $\Delta a \sim 0.5$~AU, $i \sim 
20^\circ$ and using a planetesimal density of $\rho = 3000$~kg~m$^{-3}$ and encounter velocity $v\sim e v_K \sim 12$~km~s$^{-1}$ we 
have for a planetesimal of radius 500~km that $t > 1$~Gyr. This is longer than the estimate of 56~Myr of \cite{R13} because they 
consider a smaller annulus and a planar problem. Thus for large planetesimals, collisional grinding is most likely unimportant, though 
future studies need to verify or deny this claim. In any case, the amount of leftover material warrants a separate investigation, in 
particular if the size distribution is shallow and most mass is in large planetesimals. Its results will be discussed in a companion 
paper.\\

Another topic that was not discussed in the previous sections was the amount of mass that is placed in the asteroid belt. With our 
definition of the asteroid belt region (perihelion $q>1.6$~AU and aphelion $Q<4.5$~AU), we find that simulations with a tack at 1.5~AU 
place roughly 0.6\% of material in the asteroid belt, while this increases to 0.9\% when the tack is at 2~AU, comparable to but 
slightly in excess of the percentage reported by \cite{W11}. The main reason our simulations with a tack at 1.5~AU have a higher 
amount of mass in the asteroid belt than theirs is due to the stronger gas drag acting on planetesimals in the beginning of the 
simulations because the surface density of our disc is higher than theirs. The higher value with the tack at 2~AU is clearly a result 
of weaker sculpting of the disc by Jupiter. All of these simulations leave us with an asteroid belt whose mass is at least an order of 
magnitude higher than the current value, with the caveat that we only used planetesimals with a size of 50~km for the purpose of the 
gas drag. Larger planetesimals would have reduced the remnant mass in the asteroid belt while a smaller size would have increased it 
\citep{M15}. It has been suggested that chaotic diffusion over 4~Gyr and giant planet evolution deplete the belt by approximately 
75\% of its mass, but then the remaining amount of mass is still inconsistent with what is observed today \citep{MM10}. Then again, 
our simulations do not take collisional erosion into account, which could erode the belt even further \citep{B05}.\\

A final topic that requires discussion is the total number of planets. When taking each set of initial conditions as a whole, the 
total number of planets is 4.4 $\pm$ 1.3 for the equal mass embryo case with a tack at 1.5~AU, 3.3 $\pm$ 0.8 for the oligarchic case 
with a tack at 1.5~AU, 4.8 $\pm$ 1.3 for the equal mass embryo case with a tack at 2~AU and 4.3 $\pm$ 0.9 for the oligarchic case with 
a tack at 2~AU. All of these are consistent with four terrestrial planets at the end. Yet, most of our simulations do not produce 
a Mercury analogue. This is a result of the initial conditions that we have employed, and it is noteworthy that the formation of 
Mercury has been mostly ignored in earlier works \citep{W11,OB14,J14,JM14,JW15}. If we are to ignore it too, then the total number of 
planets we must produce is three. All models as a whole are statistically consistent with three terrestrial planets, though some 
subsets within the four models are not. Since the formation and evolution of Mercury are currently unknown, further study is needed to 
rule out whether the oligarchic model with a tack at 2~AU can be made consistent with the current inner solar system. It is possible 
that by the end of our simulations the final system is not stable in the long-term and a very late collision could remove another 
planet. This is inconsistent with the historical evolution of the Solar System so we have to take the number of planets at the end of 
the simulations as final.

\section{Conclusions}
\label{sec:conc}
We have investigated the dynamical formation of the terrestrial planets in the framework of the Grand Tack scenario. It has been 
claimed that the Grand Tack reproduces several observed features of the inner Solar System that previous models failed to do, such as 
the low mass of Mars and the compositional gradient in the asteroid belt. We examined this scenario in more detail here but applied a 
different disc profile, a realistic mass-radius relationship, and took into account type 1 migration. We have stated our reasons 
for doing so in Section~2, and performed sensitivity tests to determine whether any of these differences matter. The answer appears 
to be 'yes': with the initial conditions and disc and migration model that we employed we statistically ruled out a tack at 1.5~AU 
because we produced an excess of Venus analogues and a deficit of Earth analogues. We attribute this excess of Venus analogues to our 
disc model because it has a higher surface density than the one of \cite{W11}. Additionally, with more than 95\% confidence, the 
semi-major axis of the most massive planet is inconsistent with Earth's location, while upon a visual inspection of their results the 
same is not true in the simulations of \cite{JM14}. Thus the inclusion of type 1 migration, and a much higher initial disc surface 
density, together with smaller radii of the planets, serve to shift the mass distribution closer to the Sun. This calls for a more 
distant tack location. We find that the model that best matches the current architecture of the terrestrial planets has a tack at 2~AU 
and oligarchic initial conditions.\\

{\footnotesize We thank Kevin Walsh for making available to us his version of SyMBA that incorporates the migration of the gas 
giants and the gas profile, Hal Levison for indicating we should include a realistic mass-radius relationship, and an anonymous 
reviewer for constructive comments. RB is grateful for financial support from the Astrobiology Center Project of the National 
Institutes of Natural Sciences (NINS) grant number AB271017 and to the Daiwa Anglo-Japanese Foundation for a Small Grant. RB and SJM 
acknowledge the John Templeton Foundation - Ffame Origins program in the support of CRiO. SJM is grateful for support by the NASA 
Exobiology Program (NNH14ZDA001N-EXO). SCW is supported by the Research Council of Norway (235058/F20 CRATER CLOCK) and through the 
Centres of Excellence funding scheme, project number 223272 (CEED). Numerical simulations were in part carried out on the PC cluster 
at the Center for Computational Astrophysics, National Astronomical Observatory of Japan.}

Abramov, O., and Mojzsis, S.J. (2016). Earth and Planetary Science Letters. -in press-\\
Levison, H.~F., Kretke, K.~A., Walsh, K.~J. \& Bottke, W.~F. \ 2015, PNAS 112, 14181 \\
Safronov, V.~S.\ 1969, Evolution of the protoplanetary cloud and formation of the earth and planets.
Translated from Russian (1969).~Jerusalem (Israel): Israel Program for Scientific Translations, Keter Publishing House, 212 p. 1

\clearpage
\begin{thebibliography}{}
\bibitem[Abramov et al.(2013)]{AB13} 
     Abramov, O., Kring, D.~A., \& Mojzsis, S.~J.\ 2013, Chemie der Erde / Geochemistry, 73, 227 
\bibitem[Albar\`{e}de et al.(2013)]{A13} 
     Albarede, F., Ballhaus, C., Blichert-Toft, J., et al.\ 2013, \icarus, 222, 44
\bibitem[D'Angelo \& Marzari(2012)]{AM12} 
     D'Angelo, G., \& Marzari, F.\ 2012, \apj, 757, 50 
\bibitem[Avice \& Marty(2014)]{AM14} 
     Avice, G., \& Marty, B.\ 2014, Royal Society of London Philosophical Transactions Series A, 372, 30260 
\bibitem[Bitsch et al.(2015)]{B14} 
     Bitsch, B., Johansen, A., Lambrechts, M., \& Morbidelli, A.\ 2015, \aap, 575, A28 
\bibitem[Bottke et al.(2005)]{B05} 
     Bottke, W.~F., Durda, D.~D., Nesvorn{\'y}, D., et al.\ 2005, \icarus, 179, 63      
\bibitem[Bottke et al.(2010)]{B10} 
     Bottke, W.~F., Walker, R.~J., Day, J.~M.~D., Nesvorny, D., \& Elkins-Tanton, L.\ 2010, Science, 330, 1527
\bibitem[Bottke et al.(2015)]{B15} 
     Bottke, W.~F.,Vokrouhlick{\'y}, D., Marchi, S., et al.\ 2015, Science, 348, 321 
\bibitem[Brasser et al.(2007)]{B07} 
     Brasser, R., Duncan, M.~J., \& Levison, H.~F.\ 2007, \icarus, 191, 413
\bibitem[Brasser et al.(2013)]{B13} 
     Brasser, R., Walsh, K.~J., \& Nesvorn{\'y}, D.\ 2013, \mnras, 433, 3417
\bibitem[Carlson et al.(2015)]{C15}
     Carlson, R.~W., Borg, L.~E., Gaffney, A.~M. \& Boyet, M. \ 2015, Philosophical Transactions of the Royal Society 372, 1
\bibitem[Chambers(2001)]{C01} 
     Chambers, J.~E.\ 2001, \icarus, 152, 205 
\bibitem[Chambers(2006)]{C06} 
     Chambers, J.\ 2006, \icarus, 180, 496
\bibitem[Cresswell \& Nelson(2008)]{cn08} 
     Cresswell, P., \& Nelson, R.~P.\ 2008, \aap, 482, 677
\bibitem[Dauphas \& Pourmand(2011)]{DP11} 
     Dauphas, N., \& Pourmand, A.\ 2011, \nat, 473, 489
\bibitem[Debaille et al.(2007)]{D07} 
     Debaille, V., Brandon, A.~D., Yin, Q.~Z., \& Jacobsen, B.\ 2007, \nat, 450, 525 
\bibitem[DeMeo \& Carry(2014)]{DC14} 
     DeMeo, F.~E., \& Carry, B.\ 2014, \nat, 505, 629 
\bibitem[Duncan et al.(1998)]{dll98}
     Duncan, M. J., Levison, H. F., \& Lee, M. H. 1998, \aj, 116, 2067
\bibitem[Elkins-Tanton et al.(2005)]{ET05} 
     Elkins-Tanton, L.~T., Hess, P.~C.,\& Parmentier, E.~M.\ 2005, Journal of Geophysical Research (Planets), 110, E12S01
\bibitem[Fang \& Margot(2013)]{FM13} 
     Fang, J., \& Margot, J.-L.\ 2013, \apj, 767, 115 
\bibitem[Guillot \& Hueso(2006)]{GH06} 
     Guillot, T., \& Hueso, R.\ 2006, \mnras, 367, L47
\bibitem[Halliday(2008)]{H08} 
     Halliday, A.~N.\ 2008, Royal Society of London Philosophical Transactions Series A, 366, 4163
\bibitem[Hansen(2009)]{H09}
     Hansen, B.~M.~S.\ 2009, \apj, 703, 1131 
\bibitem[Hartmann et al.(1998)]{H98} 
     Hartmann, L., Calvet, N., Gullbring, E., \& D'Alessio, P.\ 1998, \apj, 495, 385
\bibitem[Hayashi(1981)]{H81} 
     Hayashi, C.\ 1981, Progress of Theoretical Physics Supplement, 70, 35 
\bibitem[Hueso \& Guillot(2005)]{HG05} 
     Hueso, R., \& Guillot, T.\ 2005, \aap, 442, 703
\bibitem[Humayun et al.(2013)]{H13} 
     Humayun, M., Nemchin, A., Zanda, B., et al.\ 2013, \nat, 503, 513
\bibitem[Izidoro et al.(2014)]{I14} 
     Izidoro, A.,  Haghighipour, N., Winter, O.~C., \& Tsuchida, M.\ 2014, \apj, 782, 31 
\bibitem[Izidoro et al.(2015)]{I15} 
     Izidoro, A., Raymond, S.~N., Morbidelli, A., \& Winter, O.~C.\ 2015, \mnras, 453, 3619 
\bibitem[Jacobson \& Morbidelli(2014)]{JM14} 
     Jacobson, S.~A., \& Morbidelli, A.\ 2014, Royal Society of London Philosophical Transactions Series A, 372, 0174 
\bibitem[Jacobson et al.(2014)]{J14} 
     Jacobson, S.~A.,Morbidelli, A., Raymond, S.~N., et al.\ 2014, \nat, 508, 84
\bibitem[Jacobson \& Walsh(2015)]{JW15} 
     Jacobson, S.~A., \& Walsh, K.~J.\ 2015, Earth and Terrestrial Planet Formation, in The Early Earth: Accretion and Differentiation 
     (eds J.~Badro and M.~Walter), John Wiley \& Sons, Inc, Hoboken, NJ.
\bibitem[Kinoshita et al.(1991)]{kyn}
     Kinoshita, H., Yoshida, H., \& Nakai, H. 1991, Celest. Mech. Dyn. Astron., 50, 59
\bibitem[Kleine et al.(2005)]{K05} 
     Kleine, T., Palme, H., Mezger, K., \& Halliday, A.~N.\ 2005, Science, 310, 1671
\bibitem[Kokubo \& Ida(1996)]{ki96} 
     Kokubo, E., \& Ida, S.\ 1996, \icarus, 123, 180 
\bibitem[Kokubo \& Ida(1998)]{ki98} 
     Kokubo, E., \& Ida, S.\ 1998, \icarus, 131, 171
\bibitem[Laskar(2008)]{L08} 
     Laskar, J.\ 2008, \icarus, 196, 1
\bibitem[Lin \& Papaloizou(1986)]{LP86}
     Lin, D.~N.~C., \& Papaloizou, J.\ 1986, \apj, 309, 846 
\bibitem[Masset \& Snellgrove(2001)]{ms01}
     Masset, F., \& Snellgrove, M. 2001, \mnras, 320, L55
\bibitem[Masset \& Casoli(2009)]{MC09} 
     Masset, F.~S., \& Casoli, J.\ 2009, \apj, 703, 857
\bibitem[Marty(2012)]{m12} 
     Marty, B.\ 2012, Earth and Planetary Science Letters, 313, 56
\bibitem[Matsumura et al.(2016)]{M15} 
     Matsumura, S., Brasser, R. \& Ida, S. \ 2016, \apj, in press \\
\bibitem[Minton \& Malhotra(2010)]{MM10} 
     Minton, D.~A., \& Malhotra, R.\ 2010, \icarus, 207, 744 
\bibitem[Morbidelli \& Crida(2007)]{mc07}
     Morbidelli, A., \& Crida, A. 2007, \icarus, 191, 158
\bibitem[Morbidelli et al.(2007)]{M07} 
     Morbidelli, A., Tsiganis, K., Crida, A., Levison, H.~F., \& Gomes, R.\ 2007,\aj, 134, 1790
\bibitem[Morbidelli et al.(2012)]{Morby12} 
     Morbidelli, A., Marchi, S., Bottke, W.~F., \& Kring, D.~A.\ 2012, Earth and Planetary Science Letters, 355, 144
\bibitem[Morbidelli et al.(2015)]{Morby15} 
     Morbidelli, A., Lambrechts, M., Jacobson, S., \& Bitsch, B.\ 2015, \icarus, 258, 418
\bibitem[Nimmo \& Kleine(2007)]{NK07} 
     Nimmo, F., \& Kleine, T.\ 2007, \icarus, 191, 497
\bibitem[O'Brien et al.(2006)]{OB06} 
     O'Brien, D.~P., Morbidelli, A., \& Levison, H.~F.\ 2006, \icarus, 184, 39 
\bibitem[O'Brien et al.(2014)]{OB14} 
     O'Brien, D.~P., Walsh, K.~J., Morbidelli, A., Raymond, S.~N., \& Mandell, A.~M.\ 2014, \icarus, 239, 74
\bibitem[Ogihara \& Ida(2009)]{OI09} 
     Ogihara, M., \& Ida, S.\ 2009, \apj, 699, 824
\bibitem[Paardekooper et al.(2011)]{P11} 
     Paardekooper, S.-J., Baruteau, C., \& Kley, W.\ 2011, \mnras, 410, 293
\bibitem[Pierens \& Nelson(2008)]{PN08}
     Pierens, A., \& Nelson, R.~P.\ 2008, \aap, 482, 333 
\bibitem[Pierens \& Raymond(2011)]{PR11} 
     Pierens, A., \& Raymond, S.~N.\ 2011, \aap, 533, A131 
\bibitem[Pierens et al.(2014)]{P14}
     Pierens, A., Raymond, S.~N., Nesvorny, D., \& Morbidelli, A.\ 2014, \apjl, 795, L11 
\bibitem[Raymond et al.(2009)]{R09} 
     Raymond, S.~N., O'Brien, D.~P., Morbidelli, A., \& Kaib, N.~A.\ 2009, \icarus, 203, 644 
\bibitem[Raymond et al.(2013)]{R13} 
     Raymond, S.~N.,Schlichting, H.~E., Hersant, F., \& Selsis, F.\ 2013, \icarus, 226, 671
\bibitem[Rubie et al.(2015)]{R15} 
     Rubie, D.~C., Jacobson, S.~A., Morbidelli, A., et al.\ 2015, \icarus, 248, 89
\bibitem[Seager et al.(2007)]{S07} 
     Seager, S., Kuchner, M., Hier-Majumder, C.~A., \& Militzer, B.\ 2007, \apj, 669, 1279 
\bibitem[Tanaka et al.(2002)]{tanaka02}
     Tanaka, H., Takeuchi, T., \& Ward, W. R. 2002, \apj, 565, 1257
\bibitem[Tanaka \& Ward(2004)]{TW04} 
     Tanaka, H., \& Ward, W.~R.\ 2004, \apj, 602, 388     
\bibitem[Touboul et al.(2007)]{T07} 
     Touboul, M., Kleine, T., Bourdon, B., Palme, H., \& Wieler, R.\ 2007, \nat, 450, 1206 
\bibitem[Touboul et al.(2009)]{T09}
     Touboul, M., Kleine, T., Bourdon, B., Palme, H., \& Wieler, R.\ 2009, \icarus, 199, 245
\bibitem[Tsiganis (2015)]{KT15}
     Tsiganis, K. \ 2015, \nat 528, 202
\bibitem[Walker(2009)]{W09} 
     Walker, R.~J.\ 2009, Chemie der Erde / Geochemistry, 69, 101
\bibitem[Walsh et al.(2011)]{W11}
     Walsh, K.~J., Morbidelli, A., Raymond, S.~N., O'Brien, D.~P., \& Mandell, A.~M.\ 2011, \nat, 475, 206
\bibitem[Warren \& Wasson(1979)]{WW09} 
     Warren, P.~H., \& Wasson, J.~T.\ 1979, Reviews of Geophysics and Space Physics, 17, 73 
\bibitem[Werner et al.(2014)]{W14} 
     Werner, S.~C., Ody, A., \& Poulet, F.\ 2014, Science, 343, 1343
\bibitem[Wetherill(1980)]{W80} 
     Wetherill, G.~W.\ 1980, \araa, 18, 77
\bibitem[Wetherill \& Stewart(1989)]{WS89} 
     Wetherill, G.~W., \& Stewart, G.~R.\ 1989, \icarus, 77, 330 
\bibitem[Wisdom \& Holman(1991)]{wis91}
     Wisdom, J., \& Holman, M. 1991, \aj, 102, 1528
\bibitem[Zhang \& Zhou(2010)]{ZZ10} 
     Zhang, H., \& Zhou, J.-L.\ 2010, \apj, 714, 532 
\end{thebibliography}
\end{document}